\documentclass[aps,preprint]{revtex4}%
\usepackage{amsfonts}
\usepackage{amsmath}
\usepackage{amssymb}
\usepackage{graphicx}%
\begin{document}
\preprint{ }
\title{Competition between $\pi-$coupling and FFLO modulation in SF/SF atomic
thickness bilayers\\ }
\author{S. Tollis, J. Cayssol and A. Buzdin$^{\ast}$}
\affiliation{Condensed Matter Theory Group, CPMOH, UMR 5798, Universit\'{e} Bordeaux I,
33405 Talence, France }
\affiliation{(*) also Institut Universitaire de France, Paris}
\affiliation{}


\begin{abstract}
We present the detailed theoretical study of a heterostructure comprising of
two coupled ferromagnetic superconducting layers. Our model may be also
applicable to the layered superconductors with alternating interlayer coupling
in a parallel magnetic field. It is demonstrated that such systems exhibit a
competition between the nonuniform Larkin-Ovchinnikov-Fulde-Ferrel (FFLO)
state and the $\pi$ superconducting state where the sign of the
superconducting order parameter is opposite in adjacent layers. We determine
the complete temperature-field phase diagram. In the case of low interlayer
coupling we obtain a new $\pi$ phase inserted within the FFLO phase and
located close to the usual tricritical point, whereas for strong interlayer
coupling the bilayer in the $\pi$ state reveals a very high paramagnetic limit
and the phenomenon of field-induced superconductivity.

\end{abstract}
\maketitle

\section{\bigskip Introduction}

The question of coexistence of singlet superconductivity and magnetism has
been adressed for many years. It was found that the superconducting order
parameter is destroyed by a magnetic field both via the orbital effect
\cite{ginzburg} and the paramagnetic effect.\cite{saintjames} In the usual
case of an isotropic three-dimensional (3D) superconductor under an external
magnetic field, the orbital effect prevails and leads to the well-known
temperature-field phase diagram of conventional type I or II
superconductors.\cite{abrikosovmetal} In contrast, superconductivity is
essentially suppressed by the paramagnetic effect in presence of a
ferromagnetic exchange interaction. This is also true for
quasi-two-dimensional (2D) superconductors under in-plane magnetic field and
for heavy fermions materials wherein the orbital effect is partially quenched.
In the whole paper, the magnetism is characterized by an internal exchange
field $h$ (given in energy units) which may arise either from an externally
applied magnetic field or from ferromagnetic ordering. Note that
ferromagnetism must be weak in order to avoid complete suppression of
superconductivity. This is realized in rare-earth metals or actinides in which
the indirect exchange interaction leads to Curie temperatures of a few degrees.

Superconductors with internal homogeneous exchange field $h$ exhibit a very
special behaviour. According to Chandrasekhar\cite{chandrasekhar} and
Clogston,\cite{clogston} at zero temperature uniform superconductivity should
be destroyed when the polarization energy of the free electron gas exceeds the
energy gain due to Cooper pairing in the BCS ground state. This criterion
gives the exchange field $h_{p}(T=0)=\Delta_{0}/\sqrt{2}$ where the
superconductor should undergo a first-order transition to the normal state,
$\Delta_{0\text{ }}=1.76T_{c0}$ being the zero temperature superconducting
gap. Larkin and Ovchinnikov \cite{larkin} and Fulde and Ferrell \cite{fulde}
(FFLO) predicted the existence of a nonuniform superconducting state with
higher critical exchange field $h_{3D}^{FFLO}(T=0)=0.755\Delta_{0}>h_{p}(T=0)$
and second-order transition to the normal state. This prediction was made for
3D superconductors. In quasi-2D superconductors the critical exchange field of
the FFLO state is even higher, namely $h_{2D}^{FFLO}(T=0)=\Delta_{0}%
$,\cite{bulaevskii73} while in quasi-one-dimensional systems there is no
paramagnetic limit at all.\cite{buzdin83} The appearance of the modulated FFLO
state is related to the pairing of electrons with opposite spins which do not
have the opposite momenta anymore due to the Zeeman splitting. From now on we
focus on the 2D case for which a generic temperature-exchange field phase
diagram has been established.\cite{bulaevskii73} At low field and temperature,
the ground state is characterized by a uniform superconducting order
parameter. A tricritical point, located at $h^{\ast}=1.07T_{c0}$ and $T^{\ast
}=0.56T_{c0}$, is the meeting point of three transition lines separating the
normal metal, the uniform and the nonuniform superconductors. At
$T<0.56T_{c0}$, the (low-field) uniform superconductor is separated from the
(high-field) normal metal by a narrow FFLO nonuniform superconducting phase.
In contrast, at $T>0.56T_{c0}$, the system undergoes merely a second-order
phase transition from the uniform superconductor to the normal metal when
increasing the exchange field. The nonuniform FFLO state is settled in a small
region of the phase diagram and is very sensitive to
impurities,\cite{aslamasov,buzdin87} making it difficult to observe
experimentally. Nevertheless, several evidences of the FFLO state have been
obtained recently in organic superconductors \cite{orga1,orga2} and in heavy
fermions compounds, see Martin \textit{et al.}\cite{martin} and references therein.

In the context of organic and high-$T_{c}$ superconductors, layered systems
made of conducting atomic planes have been extensively
studied.\cite{bulaevskii90} In order to investigate the interplay of
superconductivity and magnetism in such anisotropic systems,\cite{revuebuzdin}
\ Andreev \textit{et al.} considered a periodic array of alternating
ferromagnetic and superconducting 2D planes.\cite{andreev91} Solving the
corresponding Gor'kov equations, these authors established the existence of a
$\pi$ state wherein each F layer separates superconducting planes with
opposite order parameter. This is relevant for the ruthenocuprate compound
RuSr$_{2}$GdCu$_{2}$O$_{8}$ which comprises CuO$_{2}$ superconducting planes
and RuO$_{2}$ magnetic planes.\cite{maclaughlin,nachtrab04} A related system
is an isolated F/S/F trilayer which exhibits the so-called superconducting
spin-valve effect. Namely, its critical temperature is higher in the
antiparallel (AP) orientation of the layers magnetizations than in the
parallel (P) orientation both for thick layers\cite{buzdin99,baladie} and
atomic size layers.\cite{daumens,tollis} Surprisingly, in the atomic thickness
limit, the superconducting gap at zero temperature is higher for P orientation
of the magnetizations.\cite{melin1,melin2,tollis,daumens} Hence one expects a
transition from AP to P orientation by cooling the system below a finite
crossing temperature. The recent progress in molecular beam
epitaxy\cite{bozovicingenieur} enables to fabricate such F/S/F trilayer with
atomic thicknesses.

In this paper, we consider a periodic array of SF bilayers. Each bilayer is
made of two atomic planes coupled by single electron tunneling. Both exchange
fields and BCS superconducting pairing are present in each SF plane.\ The
possibility of $\chi=\pi$ phase difference between the planes inside each
bilayer is also taken into account. In the whole paper, we assume that the
coupling $\ t^{\prime}$ between successive bilayers is considerably weaker
than the intra-bilayer coupling $\ t$. Our study is performed within the
framework of the BCS theory of s-wave superconductivity. Solving exactly the
Gor'kov equations in the limit $t^{\prime}/t\rightarrow0$, we first derive the
critical temperature and the superconducting gap both for parallel (P) and
antiparallel (AP) orientation of the magnetizations. We show that the critical
temperature is higher for the AP orientation than for the P orientation
whereas it is the opposite for the zero temperature gap. We also investigate
the interlayer Josephson current in the small coupling limit: the current
increases as a function of the exchange field for AP orientation whereas it is
field-independent for the P orientation. Furthermore, we find that for low
exchange fields and high temperatures, the ground state corresponds to
identical superconducting order parameters on adjacent layers. For high enough
fields and/or low enough temperatures, the $\pi$ phase ground state is
favoured and compete with the FFLO state. For the P orientation, the full
temperature-exchange field phase diagram is constructed in the two limits of
extremely low and high coupling between the planes. As expected, for
perturbative coupling between two SF planes, the phase diagram is very close
to the quasi-2D superconductor's phase diagram. Nevertheless an important
change arises. Indeed a new $\pi-$phase is inserted inside the usual FFLO
phase close to the tricritical point. For higher tunneling coupling $t\geq
T_{c0}$, this $\pi-$ phase is pushed to low temperatures $T\leq T_{c0}^{2}/t$
and high fields $h\approx t$. In this unusual superconducting phase, the
Zeeman splitting is compensated by the bonding/antibonding energy splitting
due to single-electron tunneling between the planes.\cite{prl} As a result,
field-induced superconductivity and enhanced paramagnetic limit are realized
in this simple model. These new phenomena are encountered due to the
introduction of an additional discrete degree of freedom, here the layer index
$j$. The layer index acts as a pseudo-spin and thus enlarges the usual
spin-space for singlet pairing. This idea was introduced by Kulic and
Hofmann\cite{kulic} in the context of two-bands superconductivity for which
the pseudo-spin was the band index. Nevertheless, these authors did not
investigate the presently studied $\pi$ state.

The outline of the paper is the following. In Sec.II, we present the model,
derive the corresponding Gor'kov equations and give their exact solutions. In
Sec.III, we investigate the critical temperature, the gap and the interlayer
Josephson current in the small exchange field regime for which there are only
uniform superconducting phases. In the last two sections the
temperature-exchange field phase diagram of the bilayer is studied thoroughly.
In Sec.IV, we first construct a Ginzburg-Landau functional to determine the
transitions between the different phases in the low interlayer coupling limit.
Sec.V is devoted to the\ opposite limit of strong interlayer coupling. In
conclusion, we discuss the conditions for the observation of field-induced superconductivity.

\section{Atomic thickness SF/SF bilayer}

We consider a superconducting ferromagnetic bilayer (see Fig.1) constituted of
two superconducting atomic layers, labeled as $j=1$ and $j=2$.\ In the whole
article, we assume $t\ll E_{F}$ where $t$ is the interlayer coupling energy
and $E_{F}$ the Fermi energy. As a consequence, Cooper pairs are localized
within each plane.\cite{bulaevskii90} Each layer $j$ supports a
superconducting singlet BCS coupling with the energy gap $\Delta_{j}$ and an
internal exchange field $h_{j}$. The Hamiltonian of the system can be written
as
\begin{equation}
H=%
{\displaystyle\sum\limits_{j=1,2}}
\left[  H_{j}^{0}+H_{j}^{BCS}+\frac{1}{\left\vert \lambda\right\vert }%
\int\mathbf{d}^{2}\mathbf{r}\Delta_{j}^{2}(\mathbf{r})\right]  +H_{t}\text{,}
\label{hamiltonien}%
\end{equation}
where $\lambda$ is the attractive BCS interaction constant and $\mathbf{r}$ is
the two-dimensional coordinate within each layer. For the layer $j$ the
kinetic and Zeeman parts of the Hamiltonian are written together as%

\begin{equation}
H_{j}^{0}=\underset{\mathbf{p}}{\sum}\xi_{j\sigma\sigma^{\prime}}%
(\mathbf{p})\psi_{j\sigma}^{+}(\mathbf{p})\psi_{j\sigma^{\prime}}(\mathbf{p}),
\end{equation}
in which summation over repeated spin indexes $\sigma$ and $\sigma^{\prime}$
is implied. Creation (resp. annihilation) operator of an electron with spin
$\sigma$ and two-dimensional momentum $\mathbf{p}$ in the layer $j$ is denoted
$\psi_{j\sigma}(\mathbf{p})$ (resp. $\psi_{j\sigma}^{+}(\mathbf{p})$). The
exchange fields $h_{j}$ are assumed to be either equal ($h_{1}=h_{2}=h$) or
opposite ($h_{1}=-h_{2}=h$). As a consequence the matrix $\xi_{j\sigma
\sigma^{\prime}}$ is spin-diagonal, and the Zeeman effect manifests itself in
breaking the spin degeneracy of the electronic energy levels according to
\begin{equation}
\xi_{j\sigma\sigma^{\prime}}(\mathbf{p})=\delta_{\sigma\sigma^{\prime}}\left[
\xi(\mathbf{p})-\sigma h_{j}\right]  \text{,}%
\end{equation}
where $\xi(\mathbf{p})=\mathbf{p}^{2}/2m-E_{F}$. The s-wave singlet
superconductivity is represented by the standard mean-field Hamiltonian
\begin{equation}
H_{j}^{BCS}=%
{\displaystyle\sum\limits_{\mathbf{p}}}
\left[  \Delta_{j}^{\ast}(\mathbf{q})\psi_{j\downarrow\text{ }}(\mathbf{p}%
)\psi_{j\uparrow\text{ }}(-\mathbf{p})+h.c.\right]  \text{,}%
\end{equation}
and the layers are coupled together by the hopping Hamiltonian
\begin{equation}
H_{t}=\underset{\mathbf{p,\sigma}}{t\sum}\left[  \psi_{1\sigma}^{+}%
(\mathbf{p})\psi_{2\sigma}(\mathbf{p})+h.c.\right]  \text{ .}%
\end{equation}
In order to investigate the occurence of modulated superconducting phases
(FFLO), we choose the following spatial dependence for the superconducting
order parameter
\begin{equation}
\Delta_{1}(\mathbf{r})=\Delta e^{i\mathbf{q.r}+i\chi/2}\text{, }\Delta
_{2}(\mathbf{r})=\Delta e^{i\mathbf{q.r}-i\chi/2}\text{,}%
\end{equation}
where $\mathbf{q}$ is the FFLO modulation wave vector and $\chi$ the
superconducting phase difference between the layers.

The above model can be solved exactly using the Green functions
\begin{align}
F_{jk}^{+}(\mathbf{p},\mathbf{p}^{\prime})  &  =\left\langle \psi
_{j\downarrow}^{+}(\mathbf{p})\psi_{k\uparrow}^{+}(\mathbf{p}^{\prime
})\right\rangle =\delta(\mathbf{p+p}^{\prime})F_{jk}^{+}(\mathbf{p}%
)\text{,}\nonumber\\
G_{jk}(\mathbf{p},\mathbf{p}^{\prime})  &  =-\left\langle \psi_{j\uparrow
}(\mathbf{p})\psi_{k\uparrow}^{+}(\mathbf{p}^{\prime})\right\rangle
=\delta(\mathbf{p}-\mathbf{p}^{\prime}+\mathbf{q})G_{jk}(\mathbf{p})\text{,}%
\end{align}
where $j$ and $k$ are the layer's indexes. The brackets mean statistical
averaging over grand-canonical distribution.\cite{abrikosov}

We obtain the following Gor'kov equations in the Fourier representation:%
\begin{equation}%
\begin{pmatrix}
i\omega-\xi_{1\uparrow}(\mathbf{p+q}) & -t & \Delta_{1} & 0\\
-t & i\omega-\xi_{2\uparrow}(\mathbf{p+q}) & 0 & \Delta_{2}\\
\Delta_{1}^{\ast} & 0 & i\omega+\xi_{1\downarrow}(\mathbf{p}) & t\\
0 & \Delta_{2}^{\ast} & t & i\omega+\xi_{2\downarrow}(\mathbf{p})
\end{pmatrix}%
\begin{pmatrix}
G_{11}(\mathbf{p+q})\\
G_{21}(\mathbf{p+q})\\
F_{11}^{+}(\mathbf{p})\\
F_{21}^{+}(\mathbf{p})
\end{pmatrix}
=%
\begin{pmatrix}
1\\
0\\
0\\
0
\end{pmatrix}
\text{,} \label{equationgorkov}%
\end{equation}
where $\omega=(2n+1)\pi T$ are the fermionic Matsubara frequencies.

In quasi-2D superconductors \cite{bulaevskii73,houzet2}\ the maximal FFLO
modulation amplitude is of the order of $\left(  \xi_{0}\right)  ^{-1}$,
$\xi_{0}$\ being the typical superconducting coherence length. This means that
with a good approximation we can consider $\xi_{j\uparrow}(\mathbf{p}%
+\mathbf{q})=\xi(\mathbf{p})-h_{j}+\mathbf{v}_{F}\mathbf{.q}$, $\mathbf{v}%
_{F}$ being the Fermi velocity vector in the plane.

Solving the Gor'kov equations (\ref{equationgorkov}) yields the anomalous
Gor'kov Green function for the $j=1$ SF layer
\begin{equation}
F_{11}^{+}=\frac{\alpha_{2}\Delta_{1}^{\ast}+t^{2}\Delta_{2}^{\ast}}%
{\alpha_{1}\alpha_{2}+(2\Delta^{2}\cos\chi-\beta)t^{2}+t^{4}}\text{,}
\label{Fcroix}%
\end{equation}
where
\begin{equation}
\alpha_{j}=\Delta^{2}-\omega_{j+}\widetilde{\omega}_{j-}\text{ \ and \ }%
\beta=\omega_{1+}\omega_{2+}+\widetilde{\omega}_{1-}\widetilde{\omega}_{2-},
\end{equation}
with $\omega_{j\pm}=i\omega\pm\xi(\mathbf{p})-h_{j}$ and $\widetilde{\omega
}_{j\pm}=i\omega\pm\xi(\mathbf{p+q})-h_{j}.$ Similar equation holds for
$F_{22}^{+}.$ Note that in the case where a lattice made of \ such SF/SF
bilayers is considered, the generalized anomalous Green function is obtained
by replacing $t^{2}$ by $\left\vert t+t^{^{\prime}}e^{-ip_{z}a}\right\vert
^{2}$ in Eq.(\ref{Fcroix}), $p_{z}$\ being the projection of the momentum
$\mathbf{p}$ along the $z$ axis and $a$ the period of the lattice. As a
consequence, a finite inter-bilayer coupling $t^{\prime}$ introduces an
anisotropy in the dispersion relation which leads to a broadening of the
electronic excitation levels.\ 

In the absence of tunneling $t=0$ we retrieve from Eq.(\ref{Fcroix}) the
anomalous Green function of a quasi-2D superconductor with the exchange field
$h_{1}$
\begin{equation}
F_{11}^{+}=\frac{\Delta_{1}^{\ast}}{\alpha_{1}}=\frac{\Delta_{1}^{\ast}%
}{\Delta^{2}-(i\omega+\xi_{1\downarrow})(i\omega-\xi_{1\uparrow})}\text{.}
\label{Fcroixchampseul}%
\end{equation}
Although the dependence on momentum has been removed for simplicity, notice
that $\xi_{j\uparrow}=\xi_{j\uparrow}(\mathbf{p}+\mathbf{q})$ and
$\xi_{j\downarrow}=\xi_{j\downarrow}(\mathbf{p}).$ The set of basic equations
(\ref{equationgorkov}) must be completed by the self-consistency equation
\begin{equation}
\Delta_{j}^{\ast}=\left\vert \lambda\right\vert N(0)T\sum_{\omega}%
\int_{-\infty}^{+\infty}d\xi F_{jj}^{+}\text{.} \label{self}%
\end{equation}
\qquad Close to the critical temperature $T_{c}$ of the second-order phase
transition, the order parameters $\Delta_{j}$ are small and Eq.(\ref{self})
can be written as
\begin{equation}
\ln\frac{T_{c}}{T_{c0}}=2T_{c}\sum_{\omega>0}\int_{-\infty}^{+\infty}%
d\xi\left(  \frac{\operatorname{Re}F_{jj}^{+}}{\Delta_{j}^{\ast}}-\frac{\pi
}{\omega}\right)  \text{,} \label{selfTnonnul}%
\end{equation}
where $T_{c0}$ is the critical temperature for the 2D superconducting single
layer in the absence of exchange field, namely for $h=t=0$. At \ zero
temperature, it is convenient to write Eq.(\ref{self}) as
\begin{equation}
\ln\frac{\Delta}{\Delta_{0}}=\int_{-\infty}^{+\infty}\frac{d\omega}{2\pi}%
\int_{-\infty}^{+\infty}d\xi\left(  \frac{F_{jj}^{+}}{\Delta_{j}^{\ast}}%
-\frac{1}{\omega^{2}+\xi^{2}+\Delta^{2}}\right)  \text{,} \label{selfT=0}%
\end{equation}
where $\Delta_{0}=\Delta(T=0,h=0,t=0)$ is the superconducting order parameter
at $T=0$ \ in the absence of exchange field and interlayer coupling.

\section{Uniform superconducting states}

In this section, we investigate phases with uniform superconductivity within
each layer. We obtain the critical temperature of the second-order
superconducting ($S$) to normal metal ($N$) phase transition and the order
parameter $\Delta(T,h,t)$ as a function of the temperature, the exchange field
$h$ and the interlayer coupling $t$. We consider both parallel (P) and
antiparallel (AP) orientations of the magnetizations, the superconducting
phase difference being either $\chi=0$ or $\chi=\pi$. We also calculate the
Josephson interlayer current when the bilayer is connected to external
superconducting leads. Most of these results are obtained in the perturbative
limit of small coupling between the layers $t\ll T_{c0}$. The field is also
assumed to be sufficiently small to prevent the occurence of a spatial
modulation of the superconductivity within the planes. Study of nonuniform
phases and strong coupling $t\gg T_{c0}$ are respectively postponed to Sec. IV
and Sec. V.

\subsection{Critical temperature}

We consider the second-order phase transition between the normal metal and the
uniform BCS superconductor. Thus the order parameters $\Delta_{1}$ and
$\Delta_{2}$ are small and the anomalous Gor'kov Green function (\ref{Fcroix})
can be linearized in the following form
\begin{equation}
F_{11}^{+}=\frac{-\omega_{2+}\widetilde{\omega}_{2-}\Delta_{1}^{\ast}%
+t^{2}\Delta_{2}^{\ast}}{(\widetilde{\omega}_{1-}\widetilde{\omega}_{2-}%
-t^{2})(\omega_{1+}\omega_{2+}-t^{2})}\text{.} \label{Fcroixlinearisee}%
\end{equation}
where $\xi_{j\sigma}=\xi-\sigma h_{j}$. Similar equation may be found for
$F_{22}^{+}$. We first consider \textit{the parallel (P) orientation of the
magnetizations, }namely \textit{\ $h=h_{1}=h_{2}$}. The first possibility is
$\chi=0$ wherein the layers have the same superconducting order parameters
$\Delta_{1}=\Delta_{2}=\Delta$. In this situation the anomalous Green function
obtained from Eq.(\ref{Fcroixlinearisee}) is denoted $\left(  F_{11}%
^{+}\right)  ^{P,0}$. The identity
\begin{equation}%
{\displaystyle\int\limits_{-\infty}^{+\infty}}
d\xi\left(  F_{11}^{+}\right)  ^{P,0}=\frac{\pi\Delta^{\ast}}{\omega-ih},
\end{equation}
and Eq.(\ref{selfTnonnul}) yield the following implicit equation for the
critical temperature $T_{c}^{P,0}$
\begin{equation}
\ln\frac{T_{c}^{P,0}}{T_{c0}}=\Psi\left(  \frac{1}{2}\right)
-\operatorname{Re}\Psi\left(  \frac{1}{2}+i\frac{h}{2\pi T_{c}^{P,0}}\right)
\text{,} \label{TcP0}%
\end{equation}
where $\Psi(x)$ denotes the Euler digamma function. Therefore the interlayer
coupling disappears from the self-consistency equation and Eq.(\ref{TcP0}) is
identical to that for the 2D monolayer in a uniform exchange field: the
bilayer is equivalent to a single layer in the neighborhood of superconducting
to normal state transition.\cite{saintjames} The critical temperature of the
bilayer decreases when the exchange field $h$ increases. The equation
(\ref{TcP0}) describes the second-order phase transition between the normal
metal and the uniform superconductor which is realised only for fields smaller
than the tricritical one $h^{\ast}=1.07T_{c0}$. For larger fields,
superconductivity becomes nonuniform.

A second possibility is \textit{the P orientation with $\chi=\pi$ phase
difference between the layers}. Now the anomalous Gor'kov Green function is
denoted $\left(  F_{11}^{+}\right)  ^{P,\pi}$. Then%

\begin{equation}
\bigskip%
{\displaystyle\int\limits_{-\infty}^{+\infty}}
\left(  F_{11}^{+}\right)  ^{P,\pi}d\xi=\frac{\pi}{2}\left[  \frac
{\Delta^{\ast}}{\omega-i(h+t)}+\frac{\Delta^{\ast}}{\omega-i(h-t)}\right]  ,
\end{equation}
and the self-consistency relation Eq.(\ref{selfTnonnul}) yield a critical
temperature $T_{c}^{P,\pi}$ given by
\begin{equation}
\ln\frac{T_{c}^{P,\pi}}{T_{c0}}=\Psi\left(  \frac{1}{2}\right)  -\frac{1}{2}%
{\displaystyle\sum\limits_{a=\pm1}}
\operatorname{Re}\Psi\left(  \frac{1}{2}+i\frac{h+at}{2\pi T_{c}^{P,\pi}%
}\right)  \text{.} \label{TcPPI}%
\end{equation}
From this expression one may notice that superconductivity\ in the $\pi$ state
is destroyed by a combination of two effective exchange fields $h\pm t$. In
the small interlayer coupling limit $t\ll T_{c}$, Eq.(\ref{TcPPI}) becomes
\begin{equation}
\ln\frac{T_{c}^{P,\pi}}{T_{c0}}=\ln\frac{T_{c}^{P,0}}{T_{c0}}-\left(  \frac
{t}{2\pi T_{c}}\right)  ^{2}K_{3}\left(  \frac{h}{2\pi T_{c}^{P,\pi}}\right)
\text{,} \label{TcPPIpetitt}%
\end{equation}
where the function $K_{3}\left(  x\right)  $ is defined and represented in
Appendix A. In the regime of low magnetic fields, namely for $h/2\pi
T_{c}^{P,\pi}<h^{\ast}/2\pi T^{\ast}=0.3$, the factor $K_{3}\left(  h/2\pi
T_{c}^{P,\pi}\right)  $ is positive and thus the critical temperature is
smaller in the $\pi$ superconducting state than in the $0$ state. However the
situation may be inverted if $h/2\pi T_{c}^{P,\pi}>0.3$. Moreover, along the
critical line, the value $h^{\ast}/2\pi T^{\ast}=0.3$ corresponds to the
tricritical point, $h^{\ast}\approx1.07T_{c0}$ and $T^{\ast}\approx0.56T_{c0}%
$, where FFLO nonuniform states appear. As a consequence one expects
competition between the $\pi$ superconducting phase and FFLO phases in the
neighborhood of the tricritical point. This competition will be detailed in
Sec. IV.

Let us focus on the case of \textit{AP orientation $h=h_{1}=-h_{2}$}.
Following the same procedure as previously, the equations for the critical
temperatures $T_{c}^{AP,\chi}$ are obtained. In the limit $t\ll T_{c0}$, it reads%

\begin{equation}
\frac{T_{c}^{AP,0}-T_{c}^{P,0}}{T_{c}^{P,0}}=2\pi T_{c}^{AP,0}t^{2}%
{\displaystyle\sum\limits_{\omega>0}}
\frac{h^{2}}{\left(  h^{2}+\omega^{2}\right)  ^{2}\omega}\text{,}
\label{TcAP0petitt}%
\end{equation}
for $\chi=0$, and
\begin{equation}
\frac{T_{c}^{AP,\pi}-T_{c}^{P,0}}{T_{c}^{P,0}}=-2\pi T_{c}^{AP,\pi}t^{2}%
{\displaystyle\sum\limits_{\omega>0}}
\frac{\omega}{\left(  h^{2}+\omega^{2}\right)  ^{2}} \label{TcAPpi}%
\end{equation}
for $\chi=\pi$. From Eqs.(\ref{TcAP0petitt},\ref{TcAPpi}) the critical
temperature is clearly higher in the $0$ phase than in the $\pi$ phase.
Therefore the $0$ phase is the more stable in this region of the $\left(
T,h\right)  $ phase diagram, \textit{i.e.} in the vicinity of the critical
temperature and for low fields $h<h^{\ast}=1.07T_{c0}.$

In conclusion, the bilayer is always in the $0$ superconducting state for
temperatures close to the critical temperature, whatever the relative
orientation of magnetizations is. A spin-valve effect is also present: the
critical temperature is higher for the AP orientation than for the P
orientation of the magnetizations .

\subsection{Zero temperature superconducting gap}

For $t=0$ and low fields $h<\Delta_{0}/\sqrt{2},$ it is well-known that the
zero temperature gap $\Delta(T=0,h,t=0)=\Delta_{0}$ is
field-independent.\cite{abrikosovmetal} For small interlayer coupling $t\ll
T_{c0}$, the anomalous Green function (\ref{Fcroix}) may be expanded to the
second order in $t$ as
\[
\frac{F_{11}^{+}}{\Delta_{1}^{\ast}}=\frac{1}{\alpha_{1}}+t^{2}\frac
{\alpha_{1}e^{i\chi}-2\Delta^{2}\cos\chi+\beta}{\alpha_{1}^{2}\alpha_{2}},
\]
where the full nonlinear dependence on $\Delta$ is kept in \bigskip$\alpha
_{1}$,$\alpha_{1}$ and $\beta.$ Then self-consistency relation (\ref{selfT=0})
becomes
\begin{equation}
\ln\frac{\Delta}{\Delta_{0}}=t^{2}%
{\displaystyle\iint}
\frac{d\omega}{2\pi}d\xi\frac{(\alpha_{1}-2\Delta^{2})\cos\chi+\beta}%
{\alpha_{1}^{2}\alpha_{2}}%
\end{equation}
where $\Delta=\Delta(T=0,h,t)$ and $\chi=0$ or $\chi=\pi$. Using the preceding
equation in \textit{the P orientation} we obtain $\Delta^{P,0}=\Delta^{P,\pi
}=\Delta_{0}$, either for $0$ or $\pi$ phase difference. As a result, the
superconducting gap $\Delta(T=0,h,t)$ is not affected by a small interlayer
coupling, at least at the order of $t^{2}$. The superconducting condensation
energy gain has also been calculated and the zero state found to be more
stable than $\pi$ state.

\textit{For the AP orientation}, the superconducting gap $\Delta^{AP,0}%
=\Delta^{AP,0}(T=0,h,t)$ is given by%

\begin{equation}
\ln\frac{\Delta^{AP,0}}{\Delta_{0}}=\frac{t^{2}}{2}\left[  -\frac{1}%
{\Delta_{0}^{2}-h^{2}}+\frac{\Delta_{0}^{2}-2h^{2}}{h(\Delta_{0}^{2}%
-h^{2})^{3/2}}\arcsin\frac{h}{\Delta_{0}}\right]  \text{,} \label{GapAPzero}%
\end{equation}
for \textit{zero phase difference. }The unphysical divergence at
$h\longrightarrow\Delta_{0}$ is removed by terms of higher order in $t$.
Expression (\ref{GapAPzero}) is the main result of this paragraph and reduces
to
\begin{equation}
\ln\frac{\Delta^{AP,0}}{\Delta_{0}}=-\frac{4t^{2}h^{2}}{3\Delta_{0}^{4}},
\label{gapAPPIchampfaible}%
\end{equation}
in the small field regime $h\ll\Delta_{0}$\textit{. }Therefore in the $\chi=0$
state and for AP orientation, the order parameter is suppressed by the
exchange field in the small coupling limit. This is surprising because AP
orientation was expected to weaken the effective exchange field and thus
enhance superconducting properties. Nevertheless such a decrease of the
superconducting order parameter has already been found in a ballistic
atomic-scaled F/S/F trilayer.\cite{daumens,tollis}

\textit{For the AP orientation and }$\mathit{\pi}$\textit{ phase difference,
}the gap $\Delta^{AP,\pi}(T=0,h,t)=\Delta_{0}$ is field and coupling
independent. Moreover the energy of the $\chi=\pi$ state does not depend on
the relative orientation of the magnetizations.

To summarize, the lowest energy corresponds to the $(P,0)$ phase.\ The
$(P,\pi)$ and $(AP,\pi)$ phases are degenerate with a somewhat higher energy
than the $(P,0)$ phase. Although we have not performed the energy calculation
in the case where the magnetizations are antiparallel and the phase difference
is $0$, we believe that the highest energy corresponds to the $(AP,0)$ phase
since its order parameter is the smallest one.

\subsection{Superconducting gap versus temperature: inversion of the proximity
effect}

We now extend our investigation of the superconducting gap to finite
temperatures. In order to determine the gap $\Delta(T,h,t)$ as a function of
temperature $T$, exchange field $h$ and coupling $t$, we analyse numerically
the self-consistency relation (\ref{selfT=0}) using the exact anomalous
Gor'kov Green function (\ref{Fcroix}). The result is shown schematically in
Fig.2. \textit{For the P orientation and $\chi=0$}, the superconducting gap
$\Delta^{P,0}(T,h,t)$ is the same as the gap $\Delta(T,h,t=0)$ of a single
layer whereas \textit{for $\chi=\pi$} the superconducting gap $\Delta^{P,\pi
}(T,h,t)$ is lowered by finite interlayer coupling. \textit{For the AP
orientation and $\chi=0$ }, the gap is smaller than $\Delta(T,h,t=0)$ for
$T<T_{i}$ and larger for $T>T_{i}$ where the inversion temperature
$T_{i}=T_{i}(h)$ depends only on the exchange field in the small interlayer
coupling limit (see Fig.2 inset). This phenomenon has been called inversion of
the proximity effect.\cite{tollis,daumens} Moreover, the gap $\Delta^{P,0}$ is
larger than $\Delta^{AP,\pi}$ for all temperatures.

According to these results, one may suggest several experiments. First we
consider a bilayer with magnetizations pinned in the AP mutual orientation. By
lowering the temperature, a $0$-$\pi$ transition is expected at some
temperature $T_{\pi}$. In the small interlayer coupling limit, this
temperature $T_{\pi}(h)$ is a function of the exchange field only (see Fig.2
inset). In contrast, the $0$ state is more favorable energetically for all
temperatures in the case of magnetizations pinned in the P orientation. As
another illustration we consider samples where the relative orientation of
magnetizations is free. Then the orientation is chosen by the system to
minimize its energy. Cooling such a bilayer will result in a switching from
the AP orientation to the P orientation at the inversion temperature
$T_{i}(h)$. The same prediction was made recently in a ballistic F/S/F
trilayer.\cite{tollis,daumens}

\subsection{Josephson current at T=0}

Here we consider that the SF/SF bilayer is connected to superconducting
electrodes. In this set-up, one may impose an arbitrary superconducting phase
difference $\chi$ between the SF layers, and thus a non dissipative Josephson
current flows through the bilayer in the direction perpendicular to the
planes. This interlayer Josephson current is evaluated here in the tunneling
limit $t\ll\Delta_{0}$ and at zero temperature. Within the Green functions
formalism, the general formula for the interlayer Josephson current is%

\begin{equation}
j=\frac{2ietN_{2D}}{\hbar}%
{\displaystyle\iint}
\frac{d\omega}{2\pi}d\xi(G_{12}-G_{21}),
\end{equation}
where $N_{2D}=m/(2\pi\hbar^{2})$ is the two-dimensional density of state per
spin direction and unit surface. Solving exactly the Gor'kov equations
(\ref{equationgorkov}) leads to%

\begin{equation}
G_{21}=\frac{t(\omega_{1+}\omega_{2+}-t^{2}-\Delta_{1}^{\ast}\Delta_{2}%
)}{\alpha_{1}\alpha_{2}+(2\Delta^{2}\cos\chi-\beta)t^{2}+t^{4}}\text{.}
\label{G21}%
\end{equation}
The function $G_{12}$\ is obtained from Eq.(\ref{G21}) by permuting the layer
indexes $1\longleftrightarrow2.$ The corresponding anharmonic current-phase
relationship is given by%
\[
j=\frac{2et^{2}N_{2D}}{\hbar}%
{\displaystyle\iint}
\frac{d\omega}{2\pi}d\xi\frac{2\Delta^{2}\sin\chi}{\alpha_{1}\alpha
_{2}+(2\Delta^{2}\cos\chi-\beta)t^{2}+t^{4}}%
\]

In the tunneling limit $t<<\Delta_{0},$ the interlayer Josephson current
becomes sinusoidal as a function of the phase difference,
\begin{equation}
j=j_{0}%
{\displaystyle\iint}
\frac{d\omega}{2\pi}d\xi\frac{2\Delta^{2}\sin\chi}{(\Delta^{2}+\xi^{2}%
+(\omega+ih_{1}))(\Delta^{2}+\xi^{2}+(\omega+ih_{2}))}, \label{courant1}%
\end{equation}
where $j_{0}=2eN_{2D}t^{2}/\hbar$ . The second harmonic $\sin2\chi$ has also
been evaluated and is smaller than the first one by a factor $(t/\Delta
_{0})^{2}$. The preceding equation (\ref{courant1}) yields the current-phase
relation both for parallel (P) and antiparallel (AP) orientation of
magnetizations. \textit{In the parallel case }the critical current does not
depend on the field as already reported in other systems since $j_{P}%
=j_{0}\sin\chi$.\cite{bergeretprl2001,xiaowei} \ \textit{For the antiparallel
orientation} and to the lowest order in $t$, the current-phase relation reads
\begin{equation}
j_{AP}=j_{0}f_{1}\left(  \frac{h}{\Delta}\right)  \sin\chi,
\end{equation}
where $f_{1}(x)=$ $\arcsin x/(x\sqrt{1-x^{2}})$ . Therefore the critical
current increases with the exchange field $h$ and even diverges for
$h=\Delta=\Delta^{AP}(T=0).$ Of course this divergence is unphysical and
should disappear if all orders in $t$ were taken into account. In Fig.3, the
critical current is shown as a function of the exchange field both for P and
AP orientations. Recently, the issue of the Josephson coupling between two
clean SF layers through an insulating layer was considered using Eilenberger
equations \cite{bergeretprl2001} or Bogoliubov-de Gennes
formalism.\cite{xiaowei} Similar results as ours were obtained: the critical
current increases with $h$ only if three conditions are met: low temperature,
very weak coupling between the SF layers and AP orientation. Otherwise the
presence of an exchange interaction suppresses the Josephson current. Using
Usadel equations, Krivoruchko demonstrated that this statement holds in the
diffusive regime for which the divergence for $h=\Delta$ is remplaced by a
regular peak.\cite{krivoruchko01} \bigskip

\bigskip

\section{Phase diagram of the weakly coupled SF/SF bilayer}

From now on, we consider the SF/SF bilayer only for the parallel (P)
orientation. Hence the results obtained in the next sections may be also
applied to a superconducting bilayer in an external in-plane magnetic field.
The present section is devoted to the weak coupling regime $t<<T_{c0}.$ In
contrast to the low field restriction of Sec. III, regions of the phase
diagram with $h/(2\pi T)>0.3$ are also investigated here. Then competition
between the FFLO and $\pi$ phases is expected to take place. Particular
attention is paid to the vicinity of the tricritical point given by
$h\approx1.07T_{c0}$ and $T\approx0.56T_{c0}.$\cite{saintjames} In order to
examine this narrow region of the $(T,h)$ plane, we construct a
Ginzburg-Landau (GL) functional from the Gor'kov equations used in the
previous sections. In the past Buzdin and Kachkachi \cite{kachkachi} derived a
generalized Ginzburg-Landau\ (GL) functional for a single SF layer that
describes the FFLO superconducting state near the tricritical point. Here we
extend this functional to a SF/SF bilayer for which it is possible to have not
only FFLO modulation within the planes but also $\chi=\pi$ superconducting
phase difference between the planes. For $\chi=0$, the physics of the bilayer
is independent of the coupling and thus the Buzdin-Kachkachi GL functional is
retrieved. In contrast for $\chi=\pi$, we obtain a free energy functional
which depends on the interlayer coupling $t$ and leads to the presence of a
superconducting $\pi$ phase.

\subsection{Ginzburg-Landau free energies}

The free energy of the SF/SF bilayer in a uniform superconducting state with
$\chi=0$ ($U-0$ state) is given by (see details in Appendix B)%

\begin{equation}
\mathit{F}_{U-0}(\tilde{\Delta},\mathit{\bar{\tau}})=\mathit{\bar{\tau}%
}\left\vert \tilde{\Delta}\right\vert ^{2}-\epsilon\left\vert \tilde{\Delta
}\right\vert ^{4}+b\left\vert \tilde{\Delta}\right\vert ^{6}\text{,}
\label{freezero}%
\end{equation}
with%

\begin{equation}
\mathit{\bar{\tau}(h,T)}=\ln\frac{T}{T_{c}}-K_{1}(\tilde{h}),\epsilon
=-\frac{K_{3}(\tilde{h})}{4},b=-\frac{K_{5}(\tilde{h})}{8},
\end{equation}
where $\tilde{h}=h/2\pi T$ and $\tilde{\Delta}=\Delta/2\pi T$ are respectively
the reduced exchange field and the reduced order parameter. Here we retrieve
the well-known case of a single SF layer. For reduced exchange fields lower
than the tricritical one $\tilde{h}<\tilde{h}^{\ast}$, the transition between
the superconducting and the normal states is a second-order one since
$\epsilon<0$. The critical line is given by the equation $\mathit{\bar{\tau
}(h,T)}=0.$ For higher fields, the transition becomes a first-order one
because $\epsilon>0$ and $b>0$. As for any first-order transition, two
conditions must be fullfilled. On one hand, the free energy (\ref{freezero})
is minimized, ($\partial\mathit{F}_{U-0}/\partial\tilde{\Delta})_{\tilde
{\Delta}=\tilde{\Delta}_{1}}=0$, and on the other hand the free energies of
the superconducting and normal phases are equal, $\mathit{F}_{U-0}%
(\tilde{\Delta}_{1})=0.$ Hence in the $(T,h)$ plane the equation for this
first-order line is $\mathit{\bar{\tau}}_{1}\mathit{(h,T)=}$ $\epsilon
^{2}/(4b)$ , and the jump of the superconducting gap at the transition is
given by $\left\vert \tilde{\Delta}_{1}\right\vert ^{2}=\epsilon/(2b)$. It is
well-known that this scenario is not realized because it is replaced by a
transition between normal metal and nonuniform
superconductivity.\cite{bulaevskii73} Nevertheless, this first-order
transition provides a useful energy scale $\tilde{\Delta}_{1}$ and a reference
line in the $(T,h)$ plane that will be used to construct a universal phase
diagram, namely a $t$-independent phase diagram valid for any weakly coupled
SF/SF bilayers, see Sec. IV.C.

The SF/SF bilayer may support opposite order parameters on the layers,
superconductivity being still uniform within each SF plane. In this so-called
$U-\pi$ state, the free energy of the bilayer depends on the reduced
interlayer coupling $\tilde{t}=t/2\pi T$ according to%

\begin{align}
\mathit{F}_{U-\pi}(\tilde{\Delta},\mathit{\bar{\tau}},\tilde{t})  &  =\left(
\mathit{\bar{\tau}}-4\epsilon\tilde{t}^{2}+8b\tilde{t}^{4}\right)  \left\vert
\tilde{\Delta}\right\vert ^{2}\label{freepi}\\
&  -\left(  \epsilon-12b\tilde{t}^{2}\right)  \left\vert \tilde{\Delta
}\right\vert ^{4}+b\left\vert \tilde{\Delta}\right\vert ^{6}.
\end{align}
For low reduced fields $\tilde{h}<\tilde{h}^{\ast},$ $\mathit{F}_{U-0}%
(\tilde{\Delta},\mathit{\bar{\tau}})<\mathit{F}_{U-\pi}(\tilde{\Delta
},\mathit{\bar{\tau}},\tilde{t})$. Hence the uniform superconducting phase
with $\chi=0$ is more stable than the $\pi$ phase, as already found in the
Sec. III. Interestingly for higher reduced fields $\tilde{h}>\tilde{h}^{\ast
},$ this $\pi$ phase is in competition with FFLO nonuniform superconducting
phases having either $\chi=0$ ($FFLO-0$) or $\chi=\pi$ ($FFLO-\pi$).

According to Buzdin and Kachkachi the order parameter $\Delta(x)=\Delta\cos
qx$ \ leads to the lowest energy.\cite{kachkachi} For the $FFLO-0$ phase, the
corresponding free energy reads%

\begin{align}
\mathit{F}_{LO-0}(\tilde{\Delta},Q,\mathit{\bar{\tau}})  &  =\left(
\frac{\mathit{\bar{\tau}}}{2}\mathit{-}2\epsilon Q^{2}+6bQ^{4}\right)
\left\vert \tilde{\Delta}\right\vert ^{2}\label{FFLOZERO1}\\
&  -\left(  \frac{3}{8}\epsilon+\frac{5b}{16}Q^{2}\right)  \left\vert
\tilde{\Delta}\right\vert ^{4}+\frac{5}{16}b\left\vert \tilde{\Delta
}\right\vert ^{6}, \label{FFLOZERO}%
\end{align}
whereas for the $FFLO-\pi$ phase, the free energy depends on the interlayer
coupling $t$ in the following manner%

\begin{align}
\mathit{F}_{LO-\pi}(\tilde{\Delta},Q,\mathit{\bar{\tau}},\tilde{t})  &
=\mathit{F}_{LO-0}(\tilde{\Delta},Q,\mathit{\bar{\tau}})\nonumber\\
&  +\left(  -2\epsilon\tilde{t}^{2}+4b\tilde{t}^{4}+24b\tilde{t}^{2}%
Q^{2}\right)  \left\vert \tilde{\Delta}\right\vert ^{2}\label{FFLOPI}\\
&  +\frac{9}{2}b\tilde{t}^{2}\left\vert \tilde{\Delta}\right\vert ^{4}.
\end{align}
The notation $Q=v_{F}q/(4\sqrt{2}\pi T)$ is introduced in Appendix B.

\subsection{\bigskip Competition between FFLO-$0$ and U-$\pi$ phases}

Now we proceed to analyse the above free energies in order to determine the
critical line between the normal and the superconducting states. We will also
describe the nature of the various superconducting states and what kinds of
transitions are encountered. We focus on the vicinity of the tricritical
point. Then $\epsilon=\overline{\epsilon}(\tilde{h}-\tilde{h}^{\ast})$ is a
linear function of the exchange field with $\overline{\epsilon}>0$, whereas
$b>0$ is nearly field and temperature independent.

For high reduced fields $\tilde{h}>\tilde{h}^{\ast}$, \ it appears that the
$U-0$ and the $FFLO-\pi$ never lead to the highest critical temperature. Then
we emphasize the competition between the two remaining phases, namely $U-\pi$
and $FFLO-0.$ The minimization of the GL functional (\ref{FFLOZERO1}) leads to
a modulation wave vector given by $Q^{2}=\epsilon/(6b)$ in the limit
$\tilde{\Delta}\longrightarrow0$, \textit{i.e.} near the critical line. The
$U-\pi$ phase is more stable than the $FFLO-0$ phase under the energy
condition $\mathit{F}_{U-\pi}(\tilde{\Delta},\mathit{\bar{\tau}},\tilde
{t})<\mathit{F}_{LO-0}(\tilde{\Delta},\sqrt{\epsilon/(6b)},\mathit{\bar{\tau}%
}),$ or equivalently for%

\begin{equation}
2(3-\sqrt{3})b\tilde{t}^{2}<\epsilon<2(3+\sqrt{3})b\tilde{t}^{2}.
\end{equation}

Therefore the uniform\ $\pi$ phase is "inserted" within the usual FFLO
superconducting state. The upper and lower values of $\tilde{h}$ between which
this new $\pi$ phase is stable depend on the particular value of the coupling
$t$. It is convenient to define a dimensionless generalized coordinate $\eta=$
$\epsilon/(2b\tilde{t}^{2})$ that quantifies the "distance" from the
tricritical point along the S/N transition line. Indeed $\eta=0$ at the
tricritical point and the $\pi$ phase is settled in the region $(3-\sqrt
{3})<\eta<(3+\sqrt{3}).$ As shown on Fig. 4, going along the critical line
from low to high fields, one expects the sequence of superconducting states:
uniform in the planes with $\chi=0$ for $\eta<0$, FFLO modulation along the
planes with $\chi=0$ for $0<\eta<(3-\sqrt{3}),$ then uniform $\pi$ state for
$(3-\sqrt{3})<\eta<(3+\sqrt{3})$ and finally FFLO modulation along the planes
with $\chi=0$ for $\eta>(3+\sqrt{3})$. In all cases, the sign of the
$\tilde{\Delta}^{4}$ coefficient in the GL free energy is always positive and
thus transitions between these superconducting states and the normal metal are
second-order ones.

\subsection{\bigskip Universal ($\tau,\eta$) phase diagram}

We now construct the phase diagram around the tricritical point. Because each
value of the coupling leads to different transition lines, we introduce the
following mapping of the thermodynamic variables%

\begin{equation}
\delta=\frac{\tilde{\Delta}}{\tilde{\Delta}_{1}}\text{,}\tau=\frac
{\mathit{\bar{\tau}}}{\mathit{\bar{\tau}}_{1}}, \label{scaling}%
\end{equation}
in order to obtain a universal phase diagram valid in the small coupling
regime. This mapping makes use of the energy scale $\tilde{\Delta}_{1}$ and of
the function $\ \mathit{\bar{\tau}}_{1}$ related to the first-order transition
between the normal state and the uniform superconducting state, see Sec. IV.A.
Then the free energies for the uniform superconducting phases
Eqs.(\ref{freezero},\ref{freepi}) become%

\begin{equation}
\frac{\mathit{F}_{U-0}(\delta,\tau)}{\mathit{F}_{0}}=\tau\left\vert
\delta\right\vert ^{2}-2\left\vert \delta\right\vert ^{4}+\left\vert
\delta\right\vert ^{6}\text{,}%
\end{equation}
and%

\begin{align}
\frac{\mathit{F}_{U-\pi}(\delta,\tau,\eta)}{\mathit{F}_{0}}  &  =\left(
\tau-8\left(  \frac{1}{\eta}-\frac{1}{\eta^{2}}\right)  \right)  \left\vert
\delta\right\vert ^{2}\\
&  -2\left(  1-\frac{6}{\eta}\right)  \left\vert \delta\right\vert
^{4}+\left\vert \delta\right\vert ^{6},
\end{align}
where $\mathit{F}_{0}=\epsilon^{3}/(8b^{2}).$ First it is straightforward to
minimize $\mathit{F}_{U-0}(\delta,\tau)$ and $\mathit{F}_{U-\pi}(\delta
,\tau,\eta)$ with respect to $\delta$. Then replacing the reduced gap $\delta$
by its equilibrium value, one obtains the equilibrium energies $F_{U-0}(\tau)$
and $F_{U-\pi}(\tau,\eta)$ \ of each superconducting phase. These energies are
functions of both field and temperature via the dimensionless thermodynamical
variables $\tau$ and $\eta$.

Using the same scaling Eq.(\ref{scaling}) the free energies of the \ $FFLO-0$ phase,%

\begin{align}
\frac{\mathit{F}_{LO-0}(\delta,Q,\tau)}{F_{0}}  &  =\left(  \frac{\tau}%
{2}-\frac{8b}{\epsilon}Q^{2}-\frac{24b^{2}}{\epsilon^{2}}Q^{4}\right)
\left\vert \delta\right\vert ^{2}\nonumber\\
&  -\left(  \frac{3}{4}+\frac{5b}{8\epsilon}Q^{2}\right)  \left\vert
\delta\right\vert ^{4}+\frac{5}{16}\left\vert \delta\right\vert ^{6},
\label{energyFFLOzero}%
\end{align}
and of the $FFLO-\pi$ phase,%

\begin{align}
\frac{\mathit{F}_{LO-\pi}(\delta,Q,\tau,\eta)}{F_{0}}  &  =\frac
{\mathit{F}_{LO-0}(\delta,Q,\tau)}{F_{0}}\nonumber\\
&  +\left(  -\frac{4}{x}+\frac{4}{\eta^{2}}+\frac{48}{\eta}\frac{b}{\epsilon
}Q^{2}\right)  \left\vert \delta\right\vert ^{2}\\
&  +\frac{9}{2\eta}\left\vert \delta\right\vert ^{4}, \label{energyFFLOpi}%
\end{align}
are also obtained for the order parameter $\Delta(x)=\Delta\cos qx$. The
equilibrium energies $\mathit{F}_{LO-0}(\tau)$ and $\mathit{F}_{LO-\pi}%
(\tau,\eta)$ of the modulated phases are obtained after minimization of
Eqs.(\ref{energyFFLOzero},\ref{energyFFLOpi}) with respect to $\delta$ and $Q$
.Then Eq.(\ref{energyFFLOzero}) enables to study the second-order phase
transition between the normal metallic state and the nonuniform FFLO state$.$
Under the assumption of second-order phase transition, it is sufficient to
consider the GL free energy up to the $\delta^{2}$ order. Then the free energy
(\ref{energyFFLOzero}) is minimal for $Q^{2}=\epsilon/(6b)$. For this
particular modulation, the critical FFLO/N line is given by $\tau=4/3.$

We now consider the transition lines between the various superconducting
states obtained in the previous paragraph, in particular the $FFLO-0$/$U-\pi$,
the $FFLO-0$/$U-0$ and the $U-\pi/U-0$ transitions. Let us focus on the
transition between the uniform phases $U-0$ and $U-\pi$. Solving $F_{U-0}%
(\tau)=F_{U-\pi}(\tau,\eta)$, we obtain the critical line $\tau(\eta)$ which
corresponds to a first order $U-0/U-\pi$ phase transition.\ However, this
transition is not realized (see Fig.4) because the transition to the
\ nonuniform superconducting state occurs before.

The $FFLO-0/U-0$ transition line is obtained in a similar way. The equation
$F_{U-0}(\tau)=F_{LO-0}(\tau)$ has the solution $\tau\approx0.913$. At this
value of $\tau$, the system undergoes a first order phase transition from the
uniform state to the modulated FFLO state. Adding higher harmonics to the
order parameter $\Delta(x)=\Delta\cos qx+\Delta^{\prime}\cos3qx+...,$ gives a
more accurate evaluation, namely $\tau\approx0.859$.

Finally, the $FFLO-0/U-\pi$ transition line is obtained from $\mathit{F}%
_{LO-0}(\tau)=\mathit{F}_{U-\pi}(\tau,\eta)$ and shown in Fig.4 in the
$(\tau,\eta)$ plane. This transition is a first-order one.

Using the mapping $(T,h,t)\rightarrow(\tau,\eta)$ of the thermodynamical
variables, we have obtained a universal phase diagram Fig.4 of all weakly
coupled SF/SF bilayers in the vicinity of the tricritical point. An important
feature of this phase diagram is the presence of a superconducting $\pi
$-phase. As an example, the phase diagram has been redrawn in the $(T,h)$
plane on Fig.5 for a particular value of the coupling $t$.

\section{Phase diagram of the strongly coupled SF/SF bilayer}

Now we consider the ballistic SF/SF bilayer in the regime of strong interlayer
coupling limit $t\gg T_{c0}$ and low temperature. A very unusual $\pi
$-superconducting state is found between a lower $h_{low}^{(II)}=t-\Delta
_{0}^{2}/4t$ and an upper $h_{up}^{(II)}=t+\Delta_{0}^{2}/4t$ critical
exchange field, and below a maximal temperature of the order of $T_{c0}^{2}%
/t$. Therefore field-induced superconductivity is obtained above
$h_{low}^{(II)}$ within the BCS theory of superconductivity. The underlying
physical mechanism is the compensation of the Zeeman splitting by the energy
splitting between bonding and antibonding electronic states of the bilayer,
see Fig.6.\cite{prl} Thus the new zero temperature paramagnetic limit
$h_{up}^{(II)}=t+\Delta_{0}^{2}/4t$ \ may be tuned far above the usual one
\cite{bulaevskii73} $h=\Delta_{0}$ merely by increasing the interlayer
coupling. This compensation also occurs for small coupling, but the $\pi
$-superconducting state is then less energetically favorable than the usual
$0$-superconducting phase as demonstrated in Sec.IV. Therefore the $(T,h)$
phase diagrams are topologically distinct in the opposite limits of small
(Sec.IV) and strong (Sec.V) coupling. We first analyse the second-order
superconducting/normal phase transition in Sec.V.A. Then the first-order
transition between uniform superconductivity and the normal state is discussed
in Sec.V.B.

\subsection{\bigskip Second-order phase transition}

Here we study the second-order phase transition between the $\pi$
superconducting state and the normal metal state, as a function of the field.
We start from the the linearized anomalous Green function (\ref{Fcroix}) for
arbitrary coupling $t$ and $\pi$ superconducting phase difference,
\begin{equation}
\frac{F_{11}^{+}}{\Delta_{1}^{\ast}}=\frac{(i\omega+h+\xi)(\xi-i\omega
-h)-t^{2}}{\left[  t^{2}-(i\omega+h-\xi)^{2}\right]  .\left[  t^{2}%
-(i\omega+h+\xi)^{2}\right]  }\text{.} \label{Fcroixgrandt}%
\end{equation}
From this equation and the self-consistency relation (\ref{self}), the
critical exchange field $h$ is shown to satisfy%

\begin{equation}
\left\vert h_{c}+t+\sqrt{(h_{c}+t)^{2}-X^{2}}\right\vert .\left\vert
h_{c}-t+\sqrt{(h_{c}-t)^{2}-X^{2}}\right\vert =4h_{0}^{2}\text{ ,}
\label{HCritFFLO1}%
\end{equation}
where $X=\left\vert \mathbf{q}\right\vert v_{F}$, and $h_{0}=\Delta_{0}/2$ is
the critical exchange field for the second-order superconducting phase
transition in a two-dimensional monolayer. One must then find the value of $X$
which maximizes the critical field $h_{c}$. If \ the $\pi$ phase is assumed to
be uniform inside each plane, namely if $\mathbf{q}=\mathbf{0}$,
Eq.(\ref{HCritFFLO1}) \ merely reduces to \ $\left\vert h_{c}^{2}%
-t^{2}\right\vert =h_{0}^{2}.$ The lower and upper critical fields are
respectively given by $h_{c}=t\pm h_{0}^{2}/2t$, in the limit $t\gg h_{0}$.
Thus at zero temperature and strong enough coupling, the superconductivity
destruction follows a very special scenario. At low fields, superconductivity
is first suppressed as usual at the paramagnetic limit $h_{2D}^{FFLO}%
=\Delta_{0}$ leading to the normal metal phase. Then further increase of the
field leads to a normal to superconducting phase transition at the lower
critical field. This superconducting $\pi$ phase is finally suppressed at the
upper critical field. This is a new paramagnetic limit which may be tuned far
above the usual one merely by choosing the coupling $t$ greater than
$\Delta_{0}$. Thorough analysis of \ Eq.(\ref{HCritFFLO1}) shows that the
upper critical field is even increased by an in-plane modulation in analogy
with the case of the two-dimensional FFLO phase.\cite{bulaevskii73} The upper
critical field is maximal for the choice $X=\left\vert \mathbf{q}\right\vert
v_{F}/2=\left\vert h_{c}-t\right\vert $, and then Eq.(\ref{HCritFFLO1})
reduces to
\begin{equation}
\left\vert h_{c}-t\right\vert .\left\vert h_{c}+t+2\sqrt{h_{c}t}\right\vert
=4h_{0}^{2}, \label{Hupperlower}%
\end{equation}
that gives the upper and lower fields $h_{up,low}^{(II)}=t\pm h_{0}^{2}/t$ in
the $t\gg\Delta_{0}$ limit. Note that the period of the modulated order
parameter $\left\vert \mathbf{q}\right\vert ^{-1}=\xi_{0}(t/\Delta_{0})$ is
larger than the corresponding period in the two-dimensional FFLO phase which
coincides with the ballistic coherence length $\xi_{0}=v_{F}/\Delta_{0}.$
\cite{bulaevskii73}

Furthermore one may derive the full temperature-field phase diagram using
Eqs.(\ref{selfTnonnul},\ref{Fcroixgrandt}) and the result is shown in Fig.7.
When the temperature is increased, the lower critical field increases whereas
the upper one decreases. Along the upper (resp. lower) critical line the FFLO
modulation is lost at some temperature $T_{up}^{\ast}$ (resp. $T_{low}^{\ast}%
$). For higher temperatures a uniform $\pi$ phase ($U-\pi$) is recovered and
the temperature dependence of the critical field is given by%

\begin{equation}
\ln\frac{T}{T_{c}}=\underset{a=\pm1}{\frac{1}{2}\sum}\operatorname{Re}\left[
\Psi\left(  \frac{1}{2}\right)  -\Psi\left(  \frac{1}{2}+i\frac{h_{c}%
(T)+at}{2\pi T}\right)  \right]  , \label{HcriticalTemperature}%
\end{equation}
where $\Psi\left(  x\right)  $ is the Digamma function and $\Psi
(1/2)=-C-2\ln2\simeq-1.963$, $C$ being the Euler constant. Finally the lower
and upper critical lines merge at field $h_{c}=t$ and temperature $T_{M}=\pi
e^{-C}T_{c0}^{2}/(4t)$ in the limit $t\gg T$. Therefore the field-induced
$\pi$ superconductivity is confined to temperatures lower than $T_{M}$. The
structure of these $U-\pi$ and the $FFLO-\pi$ phases is reminiscent of the
corresponding $U-0$ and the $FFLO-0$ phases although the former are shifted to
higher fields and lower temperatures than the later.

Above results were obtained for relatively strong coupling. For lower coupling
$t\simeq\Delta_{0}$, the $U-\pi$ and the $FFLO-\pi$ phases merge continuously
into the usual $\chi=0$ phases as shown in Fig.8, and finally disappear for
$t$ slightly smaller than $\Delta_{0}$. From an experimental point of view,
one might choose a system with intermediate coupling $t$ small enough to
settle the $\pi$ phase island in an available range of temperatures but also
large enough to separate the $\pi$ phase island from the usual superconducting
phases with $\chi=0$. In the general SF multilayer case the inter-bilayer
coupling constant $t^{\prime}$ needs to be sufficiently high to prevent from
superconductivity destruction by 2D fluctuations but also sufficiently low to
preserve the effect of field-induced superconductivity.\cite{prl,koshelev}

\bigskip

\subsection{First-order phase transition}

In the following we investigate the first-order $U-\pi/N$ transition to
determine whether it is more or less favorable than the above studied
second-order transition. The zero-temperature superconducting order
parameter\ $\Delta=\Delta(t,h,T=0)$ is calculated from the self-consistency
equation (\ref{selfT=0}) for P orientation of magnetizations and $\chi=\pi$
phase difference. At zero temperature, the difference between the energy
$E_{S}$ of the superconducting state and the energy $E_{N}$ of the normal
metal state is given by \cite{abrikosov}%
\begin{equation}
E_{S}-E_{N}=%
{\displaystyle\int\limits_{0}^{\Delta}}
\delta^{2}\frac{\partial}{\partial\delta}\left[
{\displaystyle\int}
{\displaystyle\int}
\left(  \frac{F^{+}}{\delta^{\ast}}(h,t,\omega,\xi,\delta)\right)
\frac{d\omega}{2\pi}d\xi\right]  d\delta\label{energieself}%
\end{equation}

In the limit $t\rightarrow0,$ we retrieve the well-known case of the single SF
layer.\cite{saintjames,abrikosovmetal} Then the self-consistency relation
(\ref{selfT=0}) admits two branches of solutions. The lower branch
$\Delta=\sqrt{\Delta_{0}(2h-\Delta_{0})}$, labelled (2) in the inset of Fig.9,
corresponds to a positive energy cost $E_{S}-E_{N}$. Thus this superconducting
solution is never realized. The actual superconducting gap is given by the
upper horizontal branch $\ \Delta=\Delta_{0}$, (1) in the inset of Fig.9,
which corresponds to the energy difference
\begin{equation}
E_{S}-E_{N}=-\frac{\pi}{2}\left(  \Delta_{0}^{2}-2h^{2}\right)  \text{.}%
\end{equation}
Hence the superconducting phase is settled for low fields $h\leq\Delta
_{0}/\sqrt{2}$ with a field-independent order parameter $\Delta=\Delta_{0}$.
For higher fields $h>\Delta_{0}/\sqrt{2}$, the system is in the normal phase
$\Delta=0$. Finally the zero temperature gap exbibits a jump at $h=\Delta
_{0}/\sqrt{2}$ which reveals the first-order transition from the uniform
superconducting phase to the normal phase.

In the opposite limit of strong interlayer coupling, we have obtained in
Sec.V.A. that field-induced superconductivity with $\chi=\pi$ phase difference
occurs for fields close to $t$ and at low temperatures. From the
self-consistency equation (\ref{selfT=0}) one obtains several possible
solutions for the zero-temperature superconducting gap $\Delta=\Delta
(t,h,T=0)$ as a function of the exchange field $h$, see Fig.9. For relatively
low fields $h_{2D}^{FFLO}=\Delta_{0}<h<h_{-}$ and for high fields $h>h_{+}$ ,
the bilayer is in the normal phase $\Delta=0.$ The limiting fields $h_{\pm}$
are solutions of%

\begin{equation}
\left(  \frac{h_{\pm}-t}{\Delta_{0}}\right)  ^{4}\mp2\left(  \frac
{t^{2}-h_{\pm}^{2}}{\Delta_{0}^{2}}\right)  +1=0\text{. } \label{hplusmoins}%
\end{equation}
For intermediate fields ranging between $h_{-}$ and $h_{+}$ there are three
superconducting branches. Two of them, (2') and (2\textquotedblright) are
never realized owing to their energy cost $E_{S}-E_{N}>0.$ The third branch
(1') requires more detailed analysis. Namely, it is given by the equation%

\begin{equation}
\frac{\Delta^{4}}{\Delta_{0}^{4}}-2\frac{(h+t)\Delta}{\Delta_{0}^{2}%
}+1=0\text{,\qquad} \label{branch1}%
\end{equation}
and the corresponding energy cost is%

\begin{align}
E_{S}-E_{N}  &  =-\frac{\pi\Delta^{2}}{2}+\frac{\pi(h+t)^{2}}{2}\left[
1-\sqrt{1-\left(  \frac{\Delta}{h+t}\right)  ^{2}}\right] \nonumber\\
&  +\frac{\pi(h-t)^{2}}{2}. \label{energiemoyen}%
\end{align}
Analysis of these equations reveals that $E_{S}-E_{N}$ is negative for
$h_{low}^{(I)}<h<h_{up}^{(I)}$ where $h_{low}^{(I)}=\sqrt{t^{2}-\Delta_{0}%
^{2}/(2\sqrt{2})}$ and $h_{up}^{(I)}=\sqrt{t^{2}+\Delta_{0}^{2}/(2\sqrt{2})}.$
Hence the SF bilayer undergoes first-order transition at $h=h_{low}^{(I)}$ and
$h=h_{up}^{(I)}$. This scenario is quite similar than the one for $t=0$, but
with a smaller order parameter jump at the transition. Moreover there are two
first-order transitions, respectively at $h_{low}^{(I)}$ and $h_{up}^{(I)}$
instead of one at $h=\Delta_{0}/\sqrt{2}$ .

In order to generalize the above gap calculations to finite temperatures and
determine the first-order S/N transition line, we have solved numerically
together the self-consistency equation (\ref{selfTnonnul}) and the condition
$E_{S}-E_{N}=0$. The result is given in the inset of Fig.7.

Collecting results from Sec.V.A and B. we obtain the full $(T,h)$ phase
diagram for the field-induced $\pi$ superconductivity. Note that this $\pi$
superconductivity reproduces the structure of the phase diagram in quasi-2D
superconductors \cite{bulaevskii73}\ although it is shifted to higher fields
and lower temperatures.

\section{CONCLUSION}

In this paper we have studied a periodic array of SF/SF bilayers in the limit
of small coupling between the different bilayers. The corresponding Gor'kov
equations have been solved exactly, taking into account both in-plane FFLO
modulation and arbitrary superconducting phase difference between SF layers.
The superconducting state with zero phase difference is always settled in the
low field regime, $h<T_{c0}$ for parallel (P) orientation of the
magnetizations. For antiparallel (AP) orientation, the $\pi$ state
predominates at low temperatures over the $0$ state which is settled in the
neighborhood of the critical line.\ Consequently if the system is pinned in
the antiparallel orientation, we predict a transition from the usual $\chi=0$
superconducting state to the $\pi$ state by cooling.

While the critical temperature is higher for the AP orientation, the zero
temperature order parameter is larger for the P orientation. This results in a
crossing temperature $T_{i}(h)$ below which the P orientation is more suitable
for superconductivity. This temperature has been calculated as a function of
the exchange field. In an experiment where the magnetizations might be easily
reversed, one therefore expects a transition from the AP to the P orientation
by cooling the system below this crossing temperature.

In the low interlayer coupling limit, a Ginzburg-Landau functional has been
derived from the exact expression of the anomalous Gor'kov Green function. As
a main result, we have obtained a $\pi$ superconducting state located in the
vicinity of the tricritical point $(h^{\ast},T^{\ast})$. Details of the
bilayer phase diagram are obtained in this framework, including the
first-order transition lines between superconducting phases. Since increasing
the interlayer coupling enlarges the $\pi$ phase region, experimental
observation of such details of the phase diagram requires the use of SF layers
with large enough interlayer coupling, namely $t\approx0.1T_{c0}$.

Finally the case of even stronger interlayer coupling, namely $t\gg T_{c0}%
$\ has been also investigated. It appears that at low temperatures the $\pi$
superconducting state is settled for exchange fields of the order of $t$,
which are well above the Chandrasekhar-Clogston paramagnetic limit.\ Thus this
new paramagnetic limit may be tuned by varying the interlayer coupling. In the
present article we have reported the detailed structure of the phase diagram
in this regime of high magnetic field. The first-order $U-\pi/N$ transition
line is also derived.

We expect that our results may be applicable to compounds like Bi$_{2}$%
Sr$_{2}$CaCu$_{2}$O$_{8}$ under a magnetic field. Indeed such perovskite
superconductors comprise tightly coupled superconducting CuO planes separated
by BiO layers.\ However observing the field-induced superconductivity in a
reasonable range of magnetic field requires relatively low critical
temperatures which are realized in the heavily doped or underdoped regimes.
Finally the latter effect is solely related to the compensation of the energy
shift in the two layers systems by the Zeeman splitting.\ So it should be
quite general and might appear also in two band superconductors or in weakly
coupled superconducting grains.\ Note that the inhomogeneous superconductivity
has been obtained in the absence of magnetic field in two-bands
superconductors.\cite{kulic} However since the $\pi$ state was not considered
in this latter work no field-induced superconductivity had been noticed.

Bulaevskii \cite{bulaevskii73} studied thoroughly Josephson coupling in
periodic layered structures with one SF plane as unit cell. Here we have
demonstrated that systems with several SF planes as unit cell exhibit
qualitatively new phenomena like field-induced superconductivity. The simplest
case, two planes per unit cell, has been studied here. It may be regarded as a
basic approach to understand the properties of more complex ferromagnetic
superconducting compounds or artificial heterojunctions.

\bigskip

We thank M.\ Daumens, M.\ Faure, M.\ Houzet and M.\ Kulic for useful
discussions and comments. This work was supported, in part, by ESF
\textquotedblright Pi-shift\textquotedblright\ Program.

\section{\ APPENDIX}

\subsection{\bigskip\bigskip Definition and properties of the functions
$K_{\mu}(\tilde{h})$}

We define the function $K_{1}(\tilde{h})$ by%

\begin{equation}
K_{1}(\tilde{h})=%
{\displaystyle\sum\limits_{n=0}^{\infty}}
\operatorname{Re}\left[  \frac{1}{n+\frac{1}{2}+i\tilde{h}}-\frac{1}%
{n+\frac{1}{2}}\right]  ,
\end{equation}

and for any integrer $\mu\geq2$ the function $K_{\mu}(\tilde{h})$\ is given by%
\begin{equation}
K_{\mu}(\tilde{h})=%
{\displaystyle\sum\limits_{n=0}^{\infty}}
\operatorname{Re}\frac{1}{\left(  n+\frac{1}{2}+i\tilde{h}\right)  ^{\mu}%
}\text{.}%
\end{equation}

The variations of $K_{1}$, $K_{3}$, $K_{5}$, with $\tilde{h}$ are represented
in Fig.10.\ One can notice that in the vicinity of the tricritical point,
\textit{i.e.} $\tilde{h}\approx\tilde{h}^{\ast}\approx0.3$, the functions
$K_{1}(\tilde{h})$ and $K_{5}(\tilde{h})$ are negative and of the order of
unity. $K_{3}(\tilde{h})$\ cancels exactly at $\tilde{h}=\tilde{h}^{\ast}$ and
becomes negative in the domain $\tilde{h}\geq\tilde{h}^{\ast}$, which is
studied Sec.IV.

\subsection{Ginzburg-Landau functional}

This part of the Appendix refers to Sec. IV of the paper.\ In the
Ginzburg-Landau theory, the free energy is expanded in terms of the gap
$\Delta$, \textit{i.e.} the order parameter, assuming the temperature close to
$T_{c}$. Originaly it was introduced as a phenomenological theory for
superconductivity before the BCS theory. Here we derive the Ginzburg-Landau
free energy from the full microscopic knowledge of our model in order to
analyze the vicinity of the tricritical point. To do this, we consider the
simplest case where the FFLO gap modulation is exponential, namely
$\Delta(x)=\Delta e^{iqx}$, $q$ being the in-plane modulation wave vector. It
is known that this modulation structure is not realized to the benefit of the
cosine modulation discussed in the article's body. However, In Sec.II, the
Gor'kov Green functions of the SF/SF bilayer were derived for a modulated
order parameter $\Delta(x)=\Delta e^{iqx}$ and $\chi=0$ or $\pi.$ This
modulation structure is then convenient to calculate the coefficients of the
generalized GL functional because the exact expression of the anomalous Green
function (see Eq.(\ref{Fcroix})) is valid for this gap modulation structure,
whereas it is unknown with the cosine structure. We first expand the exact
anomalous Green function (\ref{Fcroix}) and the self-consistency relation in
powers of the gap $\Delta$ and the FFLO wave vector $q.$ Then this
self-consistency relation is interpreted as the stationarity condition for the
Ginzburg-Landau free energy, which allows (by identification) to determine the
coefficient of every term of the GL functional.

\bigskip In the $\chi=0$ case, the expansion of the anomalous Green function
reads\ \
\begin{equation}
\frac{F_{11}^{+}}{\Delta^{\ast}}=\frac{1}{2}%
{\displaystyle\sum\limits_{n=1}^{\infty}}
\frac{\left(  -1\right)  ^{n+1}\left\vert \Delta\right\vert ^{2n-2}}{\left(
\xi-a(t,\mathbf{q})\right)  ^{n}\left(  \xi+a(-t,\mathbf{0})\right)  ^{n}%
}+(t\leftrightarrow-t)\text{,} \label{fzeroannexe}%
\end{equation}
where \ $a(t,q)=i\omega-h+t-\mathbf{v}_{F}.\mathbf{q}$. We first consider the
case of uniform superconductivity, \textit{i.e.} $\mathbf{q=0}$. After
integration over $\xi$, we obtain:%
\begin{equation}%
{\displaystyle\int\limits_{-\infty}^{+\infty}}
\frac{d\xi}{2\pi}\frac{F_{11}^{+}}{\Delta^{\ast}}=sgn(\omega)%
{\displaystyle\sum\limits_{n=0}^{\infty}}
\left(  -1\right)  ^{n}\frac{b_{n}}{2}\frac{\left\vert \Delta\right\vert
^{2n}}{\left(  \omega+ih\right)  ^{2n+1}}\text{,} \label{gna}%
\end{equation}
with $b_{n}=\frac{(2n)!}{n!^{2}2^{2n}}=\frac{\Gamma(n+1/2)}{\Gamma(1/2)n!}$.
Note that the interlayer coupling $t$ has disappeared in Eq.(\ref{gna}). We
are now able to write down the self-consistency equation (\ref{selfTnonnul})
as an expansion in powers of $\Delta$
\begin{equation}
\ln\frac{T}{T_{c0}}-%
{\displaystyle\sum\limits_{n=0}^{\infty}}
\left(  -1\right)  ^{n}b_{n}K_{2n+1}(\tilde{h})\left\vert \tilde{\Delta
}\right\vert ^{2n}=0\text{,} \label{yo}%
\end{equation}
where the functions $K_{\mu}$ are those defined in Appendix A, and
$\tilde{\Delta}=\Delta/2\pi T$. This self-consistency relation may be
interpreted as the stationnary condition
\begin{equation}
\frac{\partial\mathit{F}_{U-0}}{\partial\tilde{\Delta}}=0 \label{minim}%
\end{equation}
for the Ginzburg-Landau free energy with uniform order parameter within each
superconducting plane and $\chi=0$ phase difference between the planes. Close
to the tricritical point, $\tilde{\Delta}$ is small and it is enough to retain
only the first term in this infinite expansion as%

\begin{equation}
\ln\frac{T}{T_{c0}}-b_{0}K_{1}(\tilde{h})+b_{1}K_{3}(\tilde{h})\left\vert
\tilde{\Delta}\right\vert ^{2}-b_{2}K_{5}(\tilde{h})\left\vert \tilde{\Delta
}\right\vert ^{4}=0\text{,}%
\end{equation}
where $\tilde{h}=h/2\pi T$. By identification with Eq.(\ref{minim}) we obtain
the GL free energy for the $U-0$ phase as a function of the variational
parameter $\tilde{\Delta}$ and the thermodynamical variable $\tilde{h}=h/(2\pi
T)$:%

\begin{align}
\mathit{F}_{U-0}(\tilde{\Delta})  &  =\left[  \ln\frac{T}{T_{c0}}-b_{0}%
K_{1}(\tilde{h})\right]  \left\vert \tilde{\Delta}\right\vert ^{2}+b_{1}%
K_{3}(\tilde{h})\frac{\left\vert \tilde{\Delta}\right\vert ^{4}}{2}-b_{2}%
K_{5}(\tilde{h})\frac{\left\vert \tilde{\Delta}\right\vert ^{6}}{3}\nonumber\\
&  =\left[  \ln\frac{T}{T_{c0}}-K_{1}(\tilde{h})\right]  \left\vert
\tilde{\Delta}\right\vert ^{2}+\frac{K_{3}(\tilde{h})}{4}\left\vert
\tilde{\Delta}\right\vert ^{4}-\frac{K_{5}(\tilde{h})}{8}\left\vert
\tilde{\Delta}\right\vert ^{6} \label{energyzero}%
\end{align}
which corresponds to Eq.(\ref{freezero}). Note that in this usual $0$ state
the same coefficients have been already reported in Ref.\cite{kachkachi}

The same procedure may be followed when the phase difference is $\pi$. The
anomalous Green function is then
\begin{equation}
\frac{F_{11}^{+}}{\Delta^{\ast}}=\frac{1}{2}%
{\displaystyle\sum\limits_{n=1}^{\infty}}
\frac{\left(  -1\right)  ^{n+1}\left\vert \Delta\right\vert ^{2n-2}}{\left(
\xi-a(t,\mathbf{q})\right)  ^{n}\left(  \xi+a(t,\mathbf{0})\right)  ^{n}%
}+(t\leftrightarrow-t), \label{fpiannexe}%
\end{equation}
and leads to the self-consistency relation which contains explicitely the
coupling $t$, via the normalized coupling $\tilde{t}=t/2\pi T$:%
\begin{equation}
\ln\frac{T}{T_{c0}}-%
{\displaystyle\sum\limits_{n=0}^{\infty}}
\left(  -1\right)  ^{n}b_{n}\left[  \frac{K_{2n+1}(\tilde{h}+\tilde
{t})+K_{2n+1}(\tilde{h}-\tilde{t})}{2}\right]  \left\vert \tilde{\Delta
}\right\vert ^{2n}=0\text{.}%
\end{equation}
From the latter expression we deduce that the coefficients of the GL free
energy for the $\chi=\pi$ state can be directly obtained using the coefficient
of $\mathit{F}_{U-0}(\tilde{\Delta})$ in which we replace $K_{2n+1}(\tilde
{h})$ by $(K_{2n+1}(\tilde{h}+\tilde{t})+K_{2n+1}(\tilde{h}-\tilde{t}))/2$.
Finally the free energy of the $U-\pi$ state is
\begin{align}
\mathit{F}_{U-\pi}(\tilde{\Delta})  &  =\left[  \ln\frac{T}{T_{c0}}%
-\frac{K_{1}(\tilde{h}+\tilde{t})+K_{1}(\tilde{h}-\tilde{t})}{2}\right]
\left\vert \tilde{\Delta}\right\vert ^{2}\nonumber\\
&  +\frac{K_{3}(\tilde{h}+\tilde{t})+K_{3}(\tilde{h}-\tilde{t})}{8}\left\vert
\tilde{\Delta}\right\vert ^{4}\nonumber\\
&  -\frac{K_{5}(\tilde{h}+\tilde{t})+K_{5}(\tilde{h}-\tilde{t})}{16}\left\vert
\tilde{\Delta}\right\vert ^{6} \label{energypi}%
\end{align}
which yields Eq.(\ref{freepi}) in the small interlayer coupling limit
$\tilde{t}\ll\tilde{h}$.

We have developped in a similar way the GL free energy in the case where the
order parameter is modulated within each superconducting plane. Using the
expressions (\ref{fzeroannexe}) and (\ref{fpiannexe}) for the anomalous Green
function of the bilayer with $\Delta(x)=\Delta e^{iqx}$ FFLO modulation
respectively in the $\chi=0$ and $\chi=\pi$ cases, one obtains the expansion
of the self-consistency equation in powers of $\Delta$ and of the FFLO wave
vector $\mathbf{q}$. Finally, after averaging over all possible orientations
of the FFLO modulation vector, the self-consistency equation reads:%
\begin{equation}
\ln\frac{T}{T_{c0}}-%
{\displaystyle\sum\limits_{n=0}^{\infty}}
{\displaystyle\sum\limits_{p=0}^{\infty}}
\left(  -1\right)  ^{n+p}c_{n,p}K_{2(n+p)+1}(\tilde{h})\left\vert
\tilde{\Delta}\right\vert ^{2n}\left(  \frac{v_{F}q}{4\pi T}\right)  ^{2p}=0
\label{yoyo}%
\end{equation}
for $\chi=0$. The coefficients $c_{n,p}$ are symmetric with respect to the
expansion indexes $n$ and $p$%
\begin{equation}
c_{n,p}=\frac{\Gamma(n+p+1/2)}{\Gamma(1/2)}\frac{\left(  n+p\right)
!}{\left(  p!\right)  ^{2}\left(  n!\right)  ^{2}}\text{.}%
\end{equation}
and related to the coefficients $b_{n}$ by $b_{n}=c_{n,0}$.

From Eq.(\ref{yoyo}) the GL free energy can be constructed using the method
described in the previous paragraph for uniform phases. We retrieve all the
coefficients already obtained by Buzdin and Kachkachi,\cite{kachkachi}
including the coefficients of the gradient terms of the generalized
functional. To derive the free energy of the $FFLO-0$ phase, we have therefore
used the BK functional with the cosine modulation which is effectively
realized in each superconducting layer. As a result, it reads
\begin{align}
\mathit{F}_{LO-0}(\tilde{\Delta},Q,\tau)  &  =\left(  \frac{\mathit{\bar{\tau
}}}{2}-2\epsilon Q^{2}+6bQ^{4}\right)  \left\vert \tilde{\Delta}\right\vert
^{2}\nonumber\\
&  -\left(  \frac{3}{8}\epsilon+\frac{5}{16}bQ^{2}\right)  \left\vert
\tilde{\Delta}\right\vert ^{4}+\frac{5}{16}b\left\vert \tilde{\Delta
}\right\vert ^{6} \label{energyFFLOzeroannexe}%
\end{align}
where $Q=v_{F}q/(4\sqrt{2}\pi T)$. In the $\chi=\pi$ state the free energy has
been derived from the BK functional in which the replacement
\begin{equation}
K_{2(n+p)+1}(\tilde{h})\longrightarrow\frac{K_{2(n+p)+1}(\tilde{h}+\tilde
{t})+K_{2(n+p)+1}(\tilde{h}-\tilde{t})}{2}%
\end{equation}
has been done in order to obtain the modified coefficients. Finally the free
energy of the $FFLO-\pi$ phase can be written as%
\begin{align}
\mathit{F}_{LO-\pi}(\tilde{\Delta},Q,\tau)  &  =\left(  \frac{\mathit{\bar
{\tau}}-4\epsilon\tilde{t}^{2}+8b\tilde{t}^{4}}{2}-2\left(  \epsilon
-12b\tilde{t}^{2}\right)  Q^{2}+6bQ^{4}\right)  \left\vert \tilde{\Delta
}\right\vert ^{2}\nonumber\\
&  -\left(  \frac{3}{8}\left(  \epsilon-12b\tilde{t}^{2}\right)  +\frac{5}%
{16}bQ^{2}\right)  \left\vert \tilde{\Delta}\right\vert ^{4}+\frac{5}%
{16}b\left\vert \tilde{\Delta}\right\vert ^{6} \label{energyFFLOpiannexe}%
\end{align}
In the article body more convenient forms of Eqs.(\ref{energyFFLOzeroannexe}%
,\ref{energyFFLOpiannexe}) involving the reduced quantities $\tau$, $\delta$
and $\eta$ are used in order to derive the universal phase diagram.

\section{Figure captions}

\bigskip

FIG.1: SF/SF bilayer.\ The interlayer coupling constant is denoted $t$. The
exchange fields $h_{j}$ can be either equal (parallel orientation) or opposite
(antiparallel orientation). The superconducting phase difference between
$\Delta_{1}$ and $\Delta_{2}$ can be either $0$ ($\Delta_{1}=\Delta_{2}$) or
$\pi$ ($\Delta_{1}=-\Delta_{2}$).

FIG.2: Schematic representation of the superconducting gap as a function of
temperature for P orientation and $\chi=0$ (thicker solid line), AP
orientation and $\chi=0$ (intermediate thickness line), and AP orientation and
$\chi=\pi$ (thiner line). All curves are given for the same value of the
exchange field that is smaller $T_{c}$ (in energy units). The inversion of the
proximity effect occurs at the temperature $T_{i}(h)$ and the transition from
$0$ state to $\pi$ state in the AP\ orientation at the temperature $T_{\pi
}(h)$. Temperatures $T_{i}(h)$ and $T_{\pi}(h)$ as a function of the field are
shown in the inset.

FIG.3: Enhancement of the critical current $j_{c}$ with the field in the AP
orientation (dashed line).\ In the P orientation $j_{c}$ does not depend on
the exchange field $h$ (solid line).

FIG.4: Universal phase diagram for weakly coupled bilayers, in $(\tau,\eta)$
coordinates. The critical line (solid line) corresponds either to a $U-\pi/N$
or to a $FFLO-0/N$ transition depending on $\eta$. The transition between the
nonuniform $FFLO-0$ superconducting phase and uniform $U-0$ (resp.$\ U-\pi$)
phase is represented with dash-dotted (resp.\ dashed) line.

FIG.5: Phase diagram in $(T,h)$ coordinates, for $t/(2\pi T_{c0})=0.07$. Only
the neighborhood of the tricritical point is represented. The lines have the
same meaning than in Fig.4. Note that the $\pi$ phase is settled in a very
narrow region of the phase diagram.

FIG.6: Excitation spectrum. Usual singlet pairing (thin line circles) between
opposite-spin electrons occupying the same orbital is affected by Zeeman
effect. In contrast, $\pi$ coupling (thick line) between two electrons
occupying a bonding and an antibonding orbitals may lead to the cancellation
of the Zeeman splitting.

FIG.7: Phase diagram for $t=3\Delta_{0}\approx5.3T_{c0}$. Thick (resp. thin)
solid lines represents second-order transition between $U-\chi$ (resp.
$FFLO-\chi$) and normal metal phase ($N$) for $\chi=0$ and $\pi$. We expect
the $U-\chi/FFLO-\chi$ transition lines (not calculated) to be in the vicinity
of the (virtual) first order $U-\chi/N$ lines (dash-dotted).

FIG.8: Phase diagram for $t=\Delta_{0}\approx1.76T_{c0}$. All lines have the
same meaning than in Fig.7.

FIG.9: Order parameter at $T=0$ in the $(P,\pi)$ state for $t=3\Delta_{0}$ as
a function of the exchange field (thick solid line).

FIG.10: Functions $K_{1}(\tilde{h})$, $K_{3}(\tilde{h})$ and $K_{5}(\tilde
{h})$ defined in Appendix B.

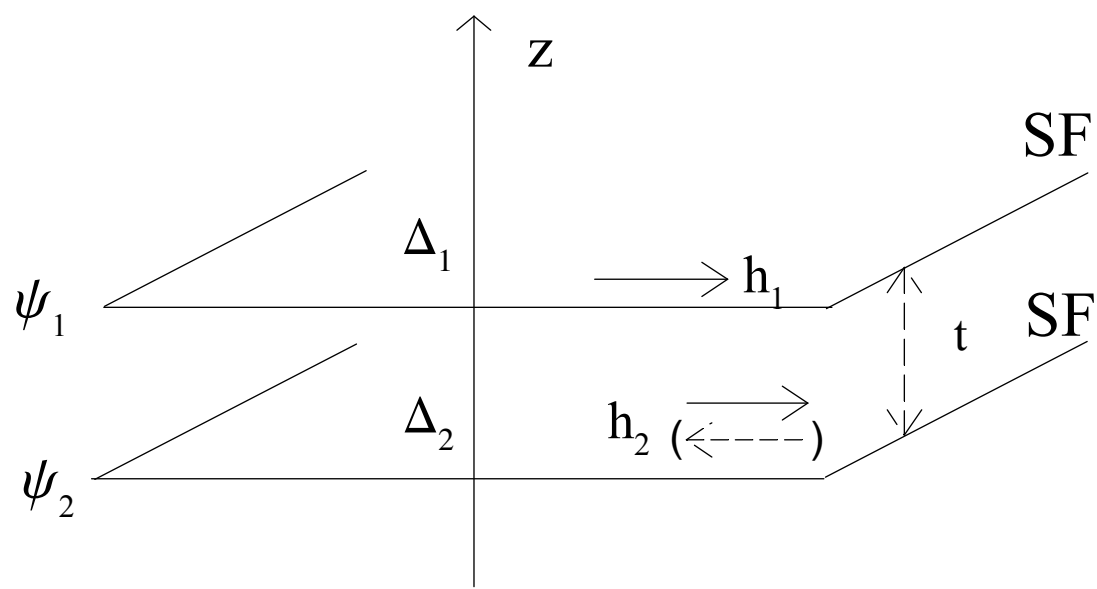
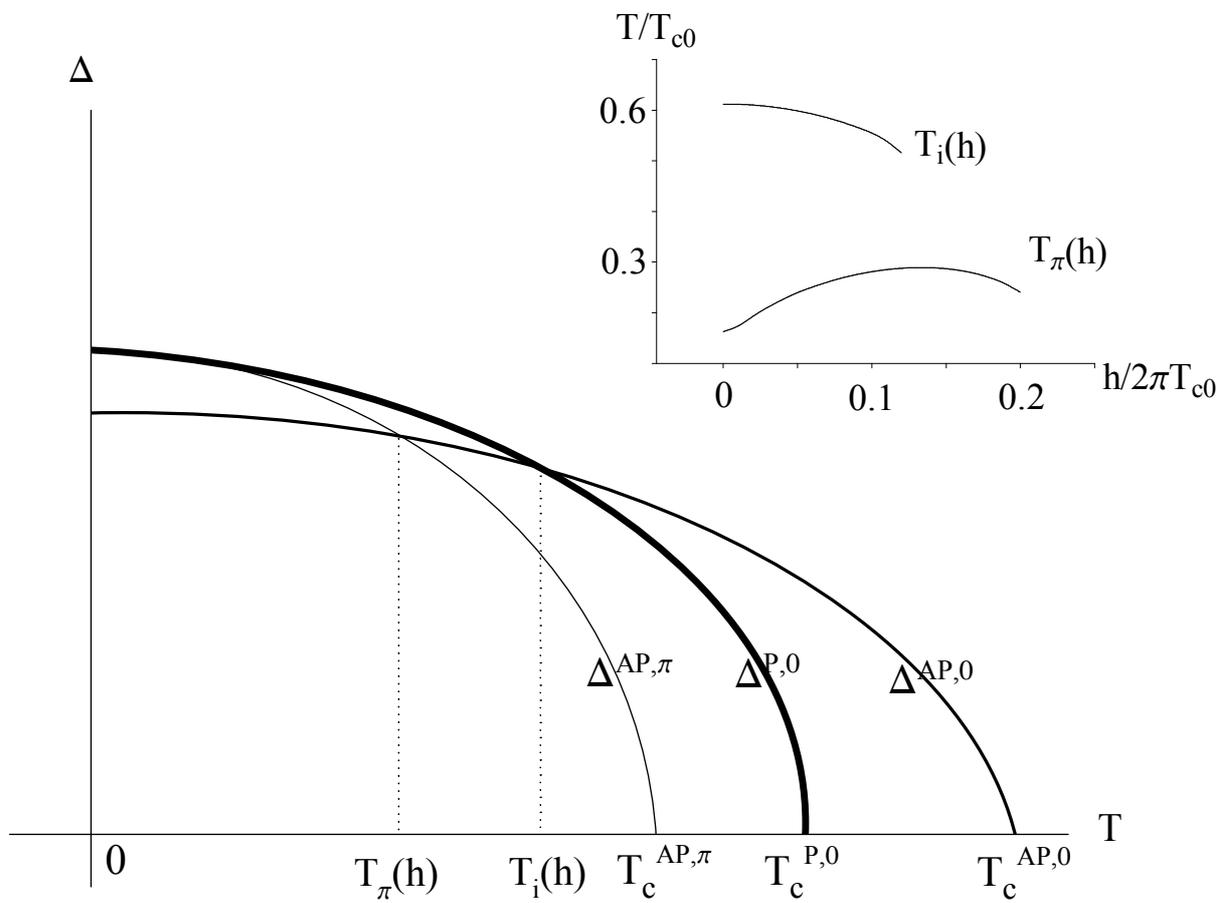
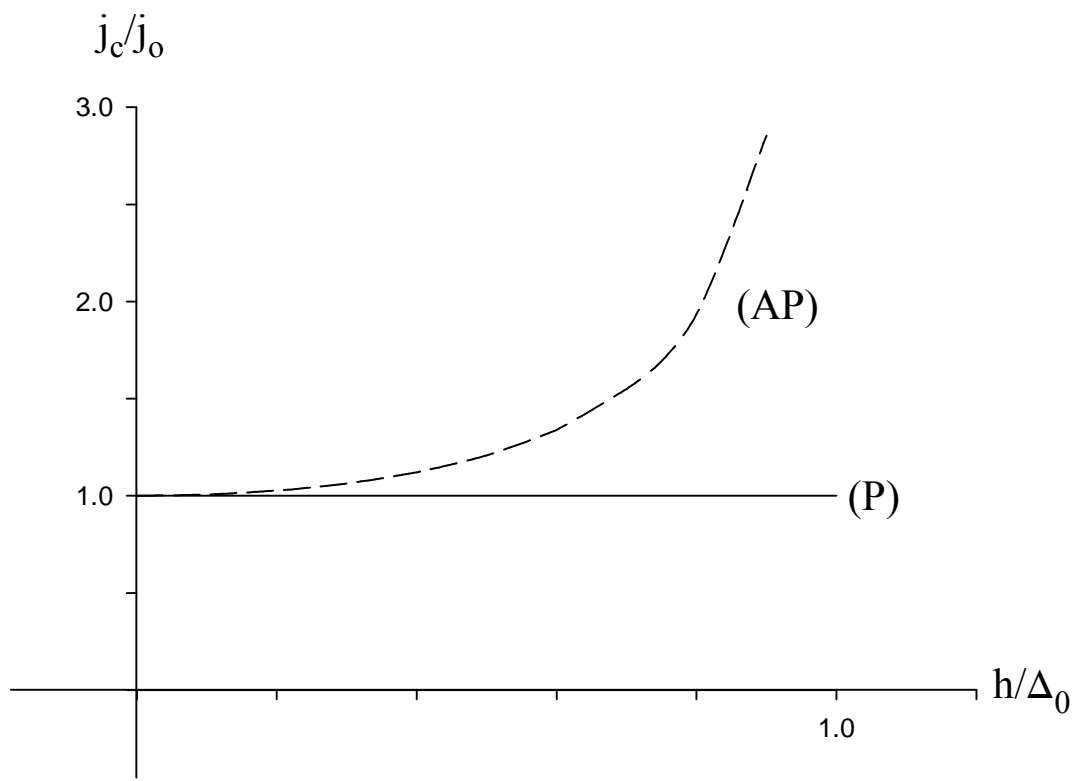
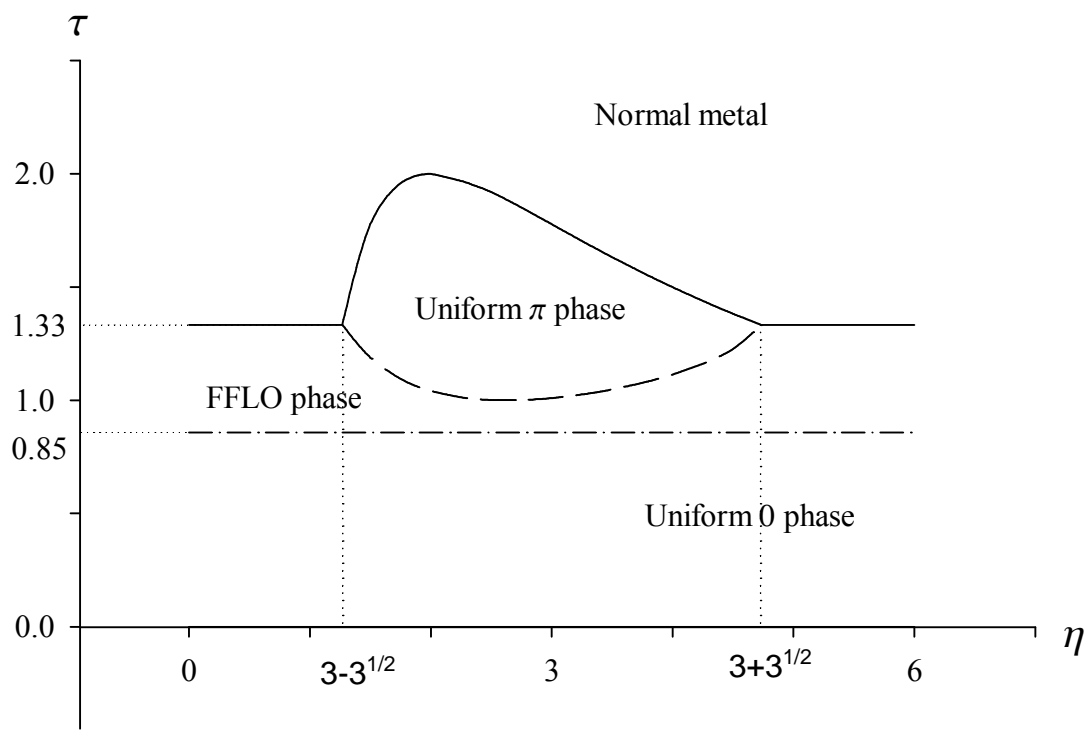
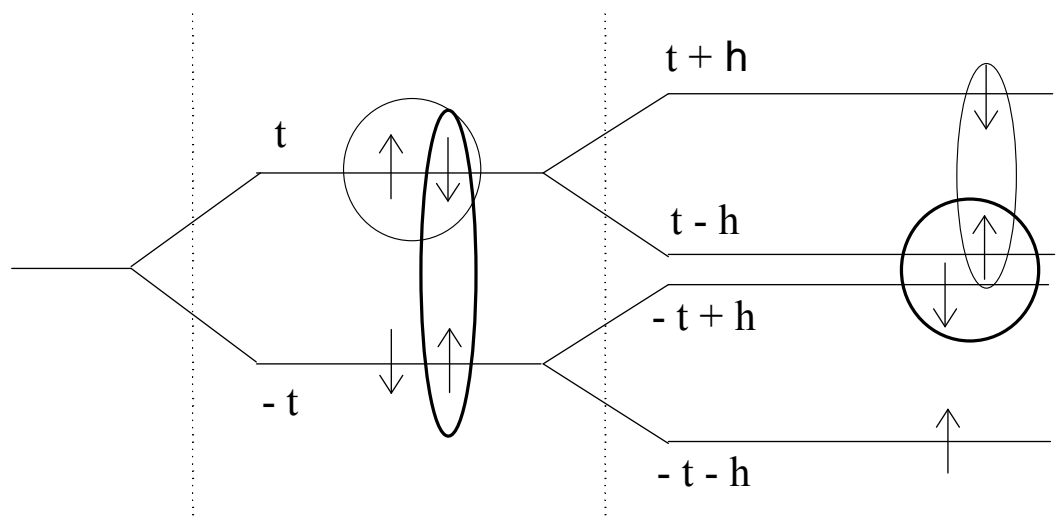
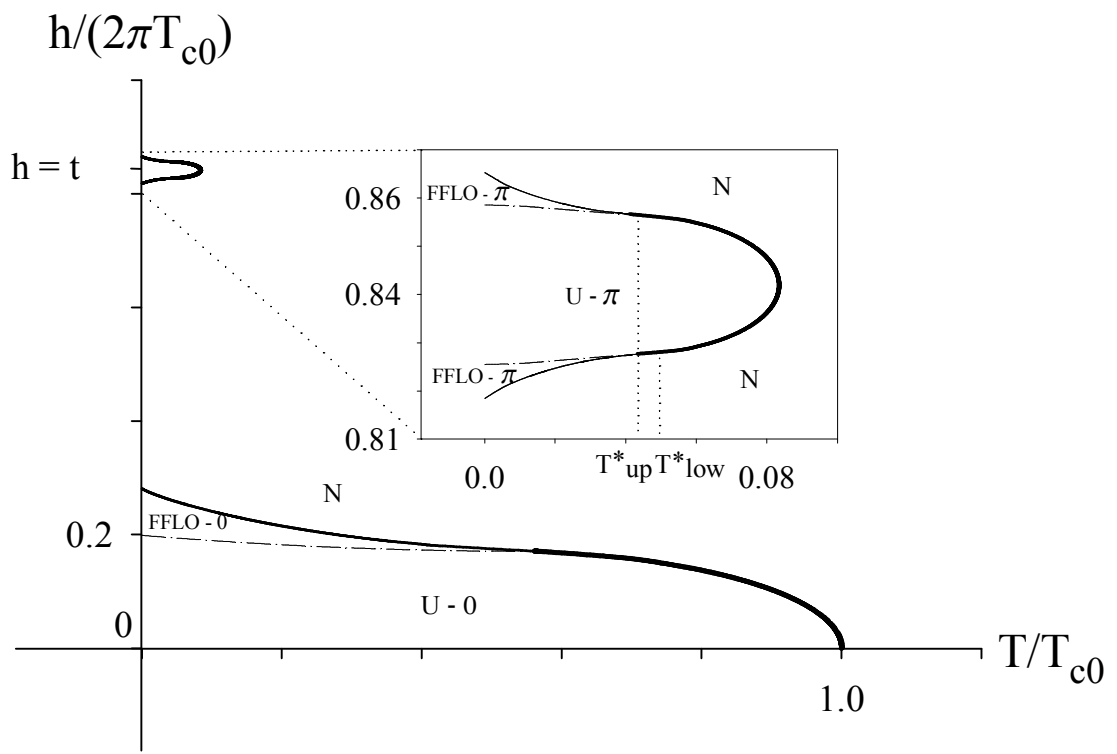
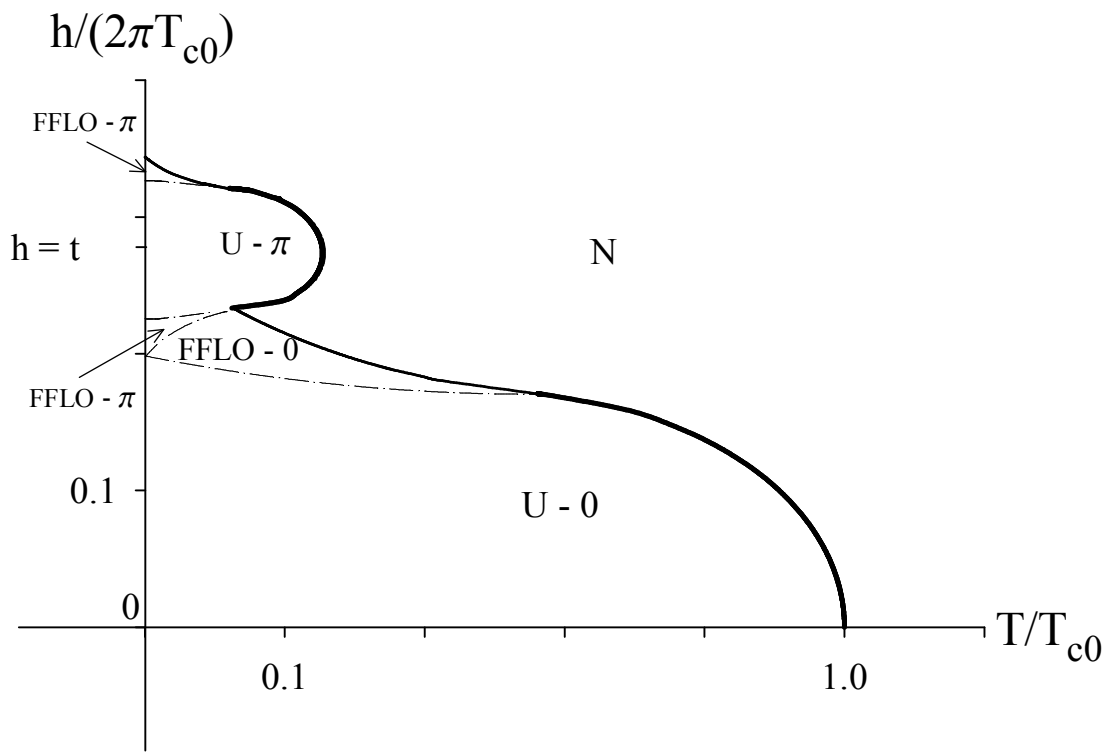
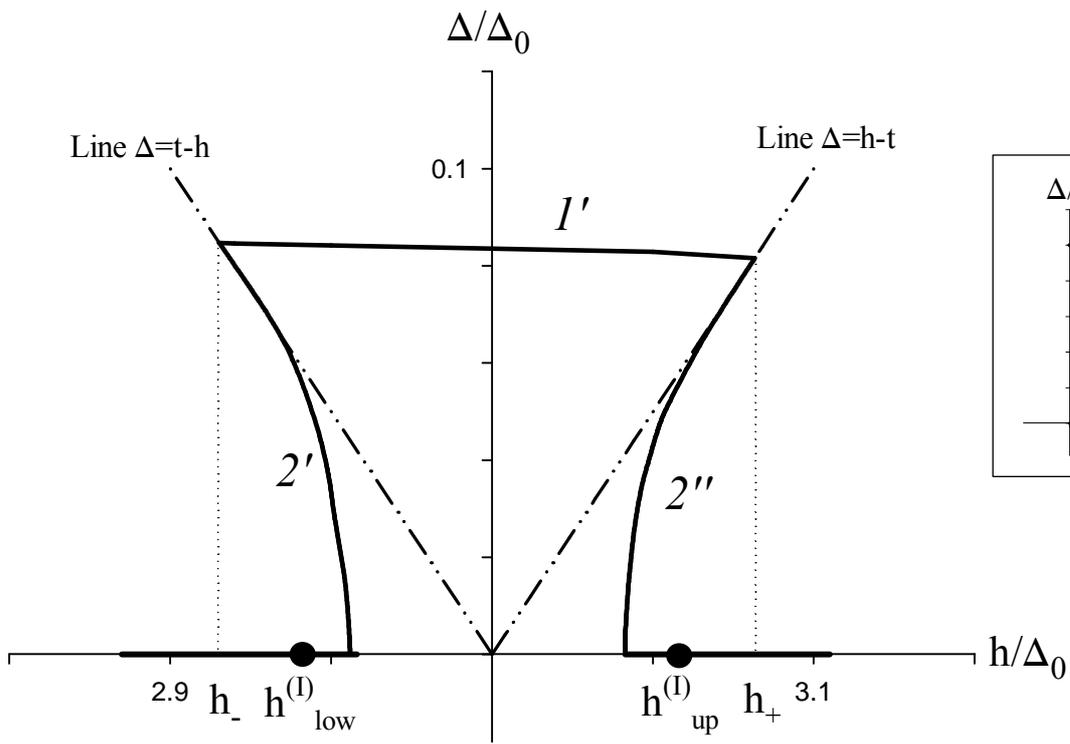
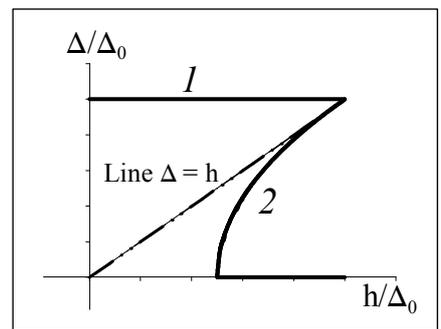
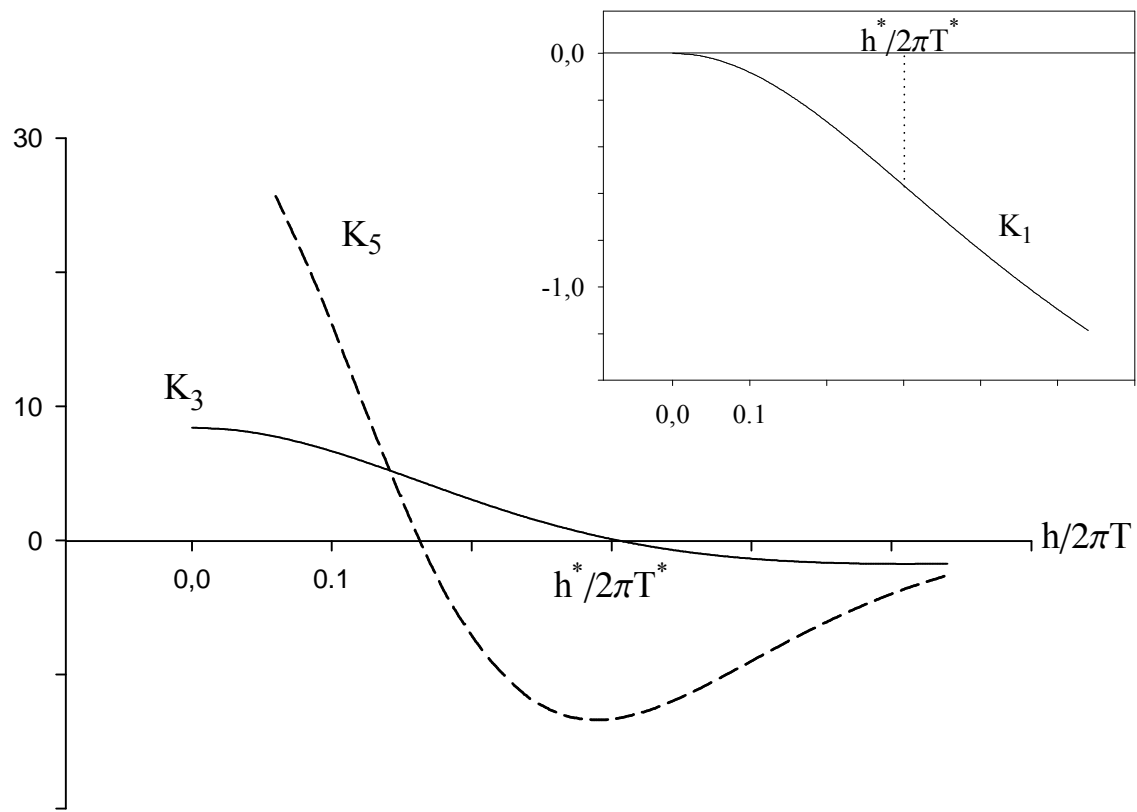
\end{document}